\documentclass[aps,pre,preprint,groupedaddress]{revtex4-1}
\usepackage{amsfonts}
\usepackage{amsmath,amssymb,amsthm}
\usepackage{eurosym}
\usepackage{graphicx}
\usepackage{dcolumn}
\usepackage{bm}
\usepackage[mathlines]{lineno}
\usepackage{epsfig}
\usepackage{epstopdf}
\usepackage{bm}
\usepackage{float}
\usepackage{xcolor}
\usepackage{graphicx}
\usepackage{subfig}
\usepackage[justification=RaggedRight]{caption}

\begin{document}

\title{Fast generation of Gaussian random fields for direct numerical simulations of stochastic transport}
\author{D. I. Palade}
\email{dragos.palade@inflpr.ro}
\author{M. Vlad}
\email{madalina.vlad@inflpr.ro}
\affiliation{ National Institute of Laser, Plasma and Radiation Physics,
		PO Box MG 36, RO-077125 M\u{a}gurele, Bucharest, Romania }


\begin{abstract}
	We propose a novel discrete method of constructing Gaussian Random Fields (GRF) based on a combination of modified spectral representations, Fourier and Blob. The method is intended for Direct Numerical Simulations of the V-Langevin equations. The latter are stereotypical descriptions of anomalous stochastic transport in various physical systems. From an Eulerian perspective, our method is designed to exhibit improved convergence rates. From a Lagrangian perspective, our method offers a pertinent description of particle trajectories in turbulent velocity fields: the exact Lagrangian invariant laws are well reproduced. From a computational perspective, our method is twice as fast as standard numerical representations.
\end{abstract}

\maketitle

\section{Introduction}

\label{section_0.0}

Stochastic phenomena are ubiquitous in nature and laboratory, being present in various sciences: physics and chemistry \cite{kampen2007stochastic}, biology \cite{bressloff2014stochastic}, finances \cite{paul2013stochastic}, social sciences \cite{diekmann2014stochastic}, etc. In particular, physical stochastic processes such as turbulent flows \cite{monin1971statistical}, anomalous transport in fusion plasmas \cite{BALESCU200762,balescu2005aspects}, flows through porous media \cite{GANAPATHYSUBRAMANIAN2009591}, seismic motion \cite{Liu2019} are complex phenomena that are modeled by nonlinear stochastic (partial) differential equations \cite{boivin_simonin_squires_1998}. Most of the theoretical studies \cite{PhysRevLett.76.4360,PhysRevE.54.1857,radivojevic2020modified,doi:10.1063/1.873745} are based on direct numerical simulations (DNSs) (or Monte Carlo simulations). Unfortunately, the ensemble statistics for the input processes as well as for the solutions exhibit slow convergence rates, with fluctuations that decay, usually, as $M^{-1/2}$, where $M$ is the dimension of the ensemble (the number of realizations). Thus, the numerical effort involved in a DNS is a matter of concern, even in the context of the computing power available nowadays \cite{yang2017direct}.

The Gaussian random fields (GRFs) \cite{abrahamsen1997a} are input stochastic processes for a large class of models. A DNS requires the generation of a large ensemble of fields, which represents an important fraction of the computing time. The topic of GRFs representation is old and a large amount of constructing techniques are available \cite{Liu2019,cuevas2020fast,solin2020hilbert,doi:10.1063/1.4789861}. From all those, by far, the most employed (especially in the context of DNSs of stochastic transport and trajectory statistics \cite{BALESCU200762,PhysRevLett.76.4360,TAUTZ20124537,boivin_simonin_squires_1998,doi:10.1063/1.2360173}) is the spectral method of discrete Fourier decomposition \cite{Liu2019} implemented using the fast Fourier transform (FFT) algorithms. This standard method exhibits some important flaws. Mathematically, the resulting fields and their covariance functions are periodic, a feature that is not usually required by the statistical model. Computationally, the convergence of the statistical properties is slow. Physically, invariants of motion may be altered due to in-between grid points interpolation.  

The aim of the present study is to derive and to analyze methods of generation of GRFs having as main criterion the minimization of the computation time. We propose a novel method (the hybrid Fourier-Blob (FB) representation) that has several advantages compared to the FFT representation. It strongly improves the convergence rates of the field statistics without imposing periodicity. These improvements reduce the computation time compared to the FFT method by as much as an order of magnitude.

The theoretical results are presented in Section \ref{section_1}. Starting from general, integral, representations of a GRF, we derive discrete variants. They are modified by introducing additional random elements, which are expected to improve the convergence. Two particular types of representations are chosen (Fourier- and Blob-like), which are shown to be canonically conjugated. 

Section \ref{section_2} contains a detailed study of the accuracy of these discrete representations. Three variants of the Fourier and Blob representation are considered. Their ability to reproduce the characteristics of the GRFs is numerically analyzed at two levels. The basic level consists of a comparison of the results on the standard Eulerian quantities (covariance and distribution functions). The second level involves the DNS of a special test-particle advection process: particle stochastic motion in two-dimensional, incompressible, time-independent velocity fields. This is a Hamiltonian process with two Lagrangian invariants. One appears in each trajectory (local invariant) and the other involves the distribution of the Lagrangian velocity (statistical invariant). They provide strong benchmarks for the numerical simulation and, implicitly, for the GRF generation method. The results on the diffusion coefficients and on the distribution of the displacements are also compared. Finally, this detailed analysis permitted to find the hybrid FB method, which appears as the fastest GRF generator able to be used in DNS studies of complex stochastic advection processes. The conclusions are summarized in Section \ref{section_3}.

\section{Theory}

\label{section_1}

We consider a real GRF $\phi (\mathbf{x})$ on a $d$ dimensional space $\phi :\mathbb{R}^{d}\rightarrow \mathbb{R}$ with zero average $\langle \phi (\mathbf{x})\rangle =0$ and covariance function  $\mathcal{E}(\mathbf{x}; \mathbf{y})=\langle \phi (\mathbf{x})\phi (\mathbf{y})\rangle $, where $\langle \cdot \rangle $ is the statistical averaging operation. This field can be generally represented through a set of parametric functions $\{F(\mathbf{x};\mathbf{s})\}$ as:

\begin{align}
\phi (\mathbf{x})& =\int d\mathbf{s}F(\mathbf{x};\mathbf{s})\zeta (\mathbf{s}%
)  \label{eq_1.1a} \\
\mathcal{E%
}(\mathbf{x};\mathbf{y}) & = \int d\mathbf{s}F(\mathbf{x};\mathbf{s})F(\mathbf{y};\mathbf{s})  \label{eq_1.1b}
\end{align}%
where $\zeta (\mathbf{s})$ is an uncorrelated random variable $\langle \zeta(\mathbf{s})\zeta (\mathbf{s}^{\prime })\rangle =\delta (\mathbf{s}-\mathbf{s}^{\prime })$. It can be easily proven that Eqs. \eqref{eq_1.1a}, \eqref{eq_1.1b} reproduce the correct covariance function $\mathcal{E}(\mathbf{x};\mathbf{y})$ while its Gaussian character is guaranteed by the Central Limit Theorem. Before discussing the nature of the parametric functions $F(\mathbf{x};\mathbf{s})$ let us address the matter of discreteness.

It is tempting to pass from the integral representation \eqref{eq_1.1a}, \eqref{eq_1.1b} to a finite and discrete form in two steps: truncate and discretize the domain of integration. As we shall see, the operator $F(\mathbf{x};\mathbf{s})$ can be, usually, safely neglected outside some finite domain in the $\mathbf{s}$ space so the truncation is justified. But using a Riemann sum $\int d\mathbf{s}\rightarrow\sum_{\mathbf{s}}$ to approximate the integrals in Eqs. \eqref{eq_1.1a}, \eqref{eq_1.1b} might not be the best approach (in the sense of errors, smoothness and convergence).

\subsection{Discrete representations}

\label{section_1.1}

Let us define in the $\mathbb{R}^d$ parametric space $\{\mathbf{s}\}$ an equidistant grid of points $\{\mathbf{s}_0\}$ with the interspacing $L$, such that each point $\mathbf{s}_0$ is centered in the hypercubic domain $\mathcal{D}(\mathbf{s}_0)$ of volume $L^d$. Accordingly, the integral over parameters can be broken as: 
\begin{equation*}
\int d\mathbf{s}\equiv \sum_{\mathbf{s}_0}\int_{\mathcal{D}(\mathbf{s}_0)} d\mathbf{s}.
\end{equation*}

Considering that $F(\mathbf{x};\mathbf{s})$ are infinitely differentiable, one can Taylor expand around a grid point $\mathbf{s}_{0}$

\begin{equation*}
F(\mathbf{x};\mathbf{s})=\sum_{n}\frac{(\mathbf{s-s_{0}})^{n}}{n!}\nabla _{\mathbf{s_{0}}}^{n}F(\mathbf{x};\mathbf{s}_{0}),
\end{equation*}
and the field \eqref{eq_1.1a} can be written

\begin{equation*}
\phi (\mathbf{x})=\sum_{\mathbf{s}_{0}}\left( \sum_{n}\frac{\hat{\alpha}_{n}}{n!}\nabla _{\mathbf{s}_{0}}^{n}\right) F(\mathbf{x};\mathbf{s}_{0}),
\end{equation*}
where the coefficients 
\begin{equation*}
\hat{\alpha}_{n}=\int_{\mathcal{D}(\mathbf{0})}d\mathbf{s}\zeta (\mathbf{s})%
\mathbf{s}^{n}
\end{equation*}
are random with zero average and correlation $\langle \hat{\alpha}_{n}\hat{\alpha}_{m}\rangle \propto L^{2d}(L/2)^{n+m+1}/(n+m+1)$. In essence, we pass from an integral (dense) representation \eqref{eq_1.1a} to a discrete one by recasting the dense character in an infinite series of random variables $\hat{\alpha}_{n}$. In the limit $L\rightarrow 0$ we can cut the series at the first order $\mathcal{O}(L^{2})$ and approximate:

\begin{equation}
\phi (\mathbf{x})\approx \sum_{\mathbf{s}_{0}}\alpha _{0}F(\mathbf{x};%
\mathbf{s}_{0}+\mathbf{\beta })  \label{fialf0}
\end{equation}%
where $\alpha _{0}$ is a Gaussian variable with $L^{d}$ variance while $\mathbf{\beta }$ a Cauchy distributed variable with the scale parameter $L/\sqrt{12}$. The representation \eqref{fialf0}\ reproduces the correlation only in an approximate manner (dependent on the magnitude of $L$). Also, the density of points 
\begin{equation}
\rho =\sum_{\mathbf{s}_{0}}\langle \delta \lbrack \mathbf{x}-(\mathbf{s}_{0}+\mathbf{\beta })]\rangle  \label{density}
\end{equation}%
is a periodic fluctuating profile around the average $\rho =1/L^{d}$ with roughly $\rho /4$ amplitude.

We propose a representation of the GRF that has the structure of \eqref{fialf0} and eliminates the above disadvantages:

\begin{equation}
\phi (\mathbf{x})\approx L^{d/2}\sum_{j}\zeta _{j}F(\mathbf{x};\mathbf{s}_{j})  \label{eq_1.2}
\end{equation}%
where the random variables $\zeta _{j}$ are uncorrelated $\langle \zeta_{j}\zeta _{i}\rangle =\delta _{i,j}$ and the points $\mathbf{s}_{j}$ are uniformly random distributed with the average density $\rho=1/L^{d}$.

\subsection{Gaussian convergence}

\label{section_1.2}

The discrete form \eqref{eq_1.2} reproduces the exact covariance function even in the limit $L\rightarrow \infty $ (a single term in the sum).
The limit $L\rightarrow 0$ (large $\rho =1/L^{d}$) is required in order to achieve the multivariate Gaussian probability distribution function $P(\{\varphi_i\};\{\mathbf{x}_{i}\})=\langle \prod_{i=1}^{k}\delta[\varphi_i-\phi(\mathbf{x}_i)]\rangle.$ Our aim is to maximize the convergence rate towards Gaussian character of the series \eqref{eq_1.2} at a fixed density. In other words, we look for representations of $\phi (\mathbf{x})$ that are "Gaussian enough" with a minimal parametric density $\rho$.

We focus for simplicity further on the one-point PDF $P(\varphi;\mathbf{x})=\langle \delta[\varphi-\phi(\mathbf{x})]\rangle$. The local rate of convergence for a sum of independent variables is bounded by the Berry-Esseen theorem \cite{doi:10.1137/S0040585X97984449} as $r<C \sigma_3/\sigma_2$. In our case: 

\begin{align*}\sigma_2^2 &=\sum_j\langle |\zeta_jF(\mathbf{x};\mathbf{s}_j)|^2\rangle\\
\sigma_3 &=max\left(\frac{\langle |\zeta_jF(\mathbf{x};\mathbf{s}_j)|^3\rangle}{\langle |\zeta_jF(\mathbf{x};\mathbf{s}_j)|^2\rangle}\right).\end{align*}

In order to reproduce the exact covariance function, $\sigma_2$ is constrained to $\sigma_2^2=1$. Thus, one can maximize $r$ under the constrain $\sigma_2=1$ and obtain through a simple functional calculus that the $\zeta_j$ variables must take randomly the $\pm 1$ values, i.e. their PDF is
\begin{equation}\label{616}
p(\zeta)=\frac{1}{2}(\delta[\zeta+1]+\delta [\zeta-1]).
\end{equation}

\subsection{Canonically conjugated representations: Fourier \& Blob case}
\label{section_1.3}

The functions $F(\mathbf{x};\mathbf{s})$ \eqref{eq_1.1b} are not unique, but are defined up to any unitary transformation $U(\mathbf{x};\mathbf{s})$:

\begin{align}
F^{\prime}(\mathbf{x};\mathbf{s})& =\int d\mathbf{s}_{1}F(\mathbf{x};%
\mathbf{s}_{1})U(\mathbf{s}_{1};\mathbf{s})  \label{eq_1.3} \\
\delta (\mathbf{s}_{1}-\mathbf{s}_{2})&=\int d\mathbf{s}U(\mathbf{s}_{1};\mathbf{s})U(\mathbf{s}_{2};\mathbf{s}),
\end{align}%
where $F^{\prime}(\mathbf{x};\mathbf{s})$ is solution of Eq. \eqref{eq_1.1b}. We note that $F(\mathbf{x};\mathbf{s})$ can be interpreted either as an operator (the "square root" of the covariance operator $\mathcal{E}(\mathbf{x};\mathbf{y})$) or as a set of parametric functions which reproduces the correlation.

A particularly important unitary transformation is the Fourier transform $U(\mathbf{s}_{1};\mathbf{s})=e^{i\mathbf{s}_{1}\mathbf{s}}$ which links two sets of \textbf{canonically conjugated representations}:

\begin{equation}
F^{\prime}(\mathbf{x};\mathbf{s})=\int d\mathbf{s}_1 F(\mathbf{x};\mathbf{s}_1)e^{i \mathbf{s}_1 \mathbf{s}}  \label{eq_1.4}
\end{equation}

A particularly  important representation is suggested by the relation \eqref{eq_1.1b} as scaled eigenvectors of the covariance operator $F(\mathbf{x};\mathbf{s})=\sqrt{\lambda (\mathbf{s})}\psi _{\mathbf{s}}(\mathbf{x})$ where $\psi _{\mathbf{s}}(\mathbf{x})$ is an eigenvector and $\lambda (\mathbf{s})$ its associated eigenvalue $\int d\mathbf{y}\mathcal{E}(\mathbf{x};\mathbf{y})\psi _{\mathbf{s}}(\mathbf{y})=\lambda (\mathbf{s})\psi _{\mathbf{s}}(\mathbf{x})$. This choice yields the Karhunen-Loeve representation \cite{681430}.

We consider the case of homogeneous GRFs, $\mathcal{E}(\mathbf{x};\mathbf{y})\equiv \mathcal{E}(\mathbf{x}-\mathbf{y})$. The natural eigenvectors for a translation invariant operator are plane waves $\psi _{\mathbf{k}}(\mathbf{x})=e^{i\mathbf{k}\mathbf{x}}$ while the corresponding eigenvalues $\lambda (\mathbf{k})=S(\mathbf{k})$ where $S(\mathbf{k})$ is the spectrum, the Fourier transform of the covariance.

Thus, using the Karhune-Loeve decomposition and searching for real-valued fields, we obtain a Fourier-like parametric function $F_F$. Choosing the transformation $U(\mathbf{a};\mathbf{k})=e^{i \mathbf{a} \mathbf{k}}$ one gets from $F_F$ the "Blob-function" $F_B$:

\begin{align*}
\zeta F_F(\mathbf{x};\mathbf{k})\equiv\sqrt{S(\mathbf{k})}sin(\mathbf{k}\mathbf{x}+\frac{\pi}{4}\zeta )\\
F_{B}(\mathbf{x};\mathbf{a})=\int d\mathbf{k}\sqrt{S(\mathbf{k})}e^{-i\mathbf{k}\mathbf{x}}e^{i\mathbf{k}\mathbf{a}}\equiv F_{B}(\mathbf{x}-\mathbf{a}).
\end{align*}%

Introducing these functions in the approximative discrete form derived \eqref{eq_1.2} we obtain \textbf{the canonically conjugated Fourier and Blob representations}

\begin{align}\label{eq_0.3a}
\phi _{F}(\mathbf{x})& \approx L_{k}^{d/2}\sum_{j=1}^{N_c}\sqrt{S(\mathbf{k}_j)}sin(\mathbf{k}_j%
\mathbf{x}+\frac{\pi}{4}\zeta_j)\\
\phi _{B}(\mathbf{x})& \approx L_{a}^{d/2}\sum_{j=1}^{N_c}\zeta _{j}F_{B}(\mathbf{x}-%
\mathbf{a}_{j})\label{eq_0.3b}
\end{align}%

which differ from the discrete Fourier decomposition (FFT) or the discrete Moving-Average methods (MA) \cite{Ravalec2000} through the use of stochastic wave numbers $\mathbf{k}_j$, blob positions $\mathbf{a}_j$ and fixed phases $\zeta_j=\pm 1.$

The series \eqref{eq_1.2} becomes finite if the functions $F(\mathbf{x};\mathbf{s})$ have a compact support in the parametric space $\mathbf{s}$. Consequently, the number of terms in the sum $N_c$ is roughly the ratio between the volume of the compact support and the chosen density of parameters $\rho$.

For the Fourier representation \eqref{eq_0.3a} the compact support is the domain in the reciprocal space $\{\mathbf{k}_i\}$ where the spectrum $S(\mathbf{k}_i)$ has non-negligible values. For the Blob representation \eqref{eq_0.3b} the compact support is the domain in the real space $\{\mathbf{a}_i\}$ where the blob function $F_B(\mathbf{x}-\mathbf{a}_i)$ has non-negligible values. Thus, these two methods require a similar number of terms $N_c$ in the sum in order to calculate a realization of the field in a point $\mathbf{x}$ with a given accuracy. 

We note that the usual discrete Fourier decomposition (with fixed grid points) usually needs larger values of $N_c$, as demonstrated in the next section.

\subsection{Advantages of the Fourier and Blob representations}

\label{section_1.6}

The FFT approach is usually considered as one of the fastest construction techniques for GRFs. It allows one to compute the values of the field $\phi(\mathbf{x})$ on a physical, equidistant grid, of dimension $N_g$ using $N_g$ equidistant wavenumbers, with a numerical complexity $\mathcal{O}(N_g\log N_g)$. Using the Fourier/Blob methods \eqref{eq_0.3a},\eqref{eq_0.3b} with random $\{\mathbf{k}_{i}\}$/$\{\mathbf{a}_{i}\}$ to compute the values on the same grid requires a computational cost which scales as $\mathcal{O}(N_g\times N_c)$ where $N_c$ is the number of parameters considered in the compact support. In general, we expect $\log N_g\ll N_c$. Even in this context, the proposed methods are particularly tempting because:

\begin{enumerate}
	\item The randomness of the parameters improves the convergence rates toward Gaussianity, such that $N_c$ is required to be only a few times larger than $\log N_g$ (especially in low dimensional spaces $d = 1,2$).
	
	\item The randomness of the parameters improves the way the parametric space is spanned, allowing for smooth convergent covariance for any number of terms in the series. FFT needs dense grids to achieve that.
	
	\item The resulting fields are not spatially periodic.
	
	\item The GRFs have a preserved structure of the equipotential lines (no interpolation needed for the field values in-between grid points).
\end{enumerate}

The disadvantages are that the random parameters must be generated at every realizations. The Blob method, requires a supplementary implementation of a nearest neighbor algorithm. The Blob method might not have always analytical Blob functions.

\section{Accuracy study and DNS tests}

\label{section_2}

The Fourier \eqref{eq_0.3a} and Blob \eqref{eq_0.3b} representations are tested in the case of a 2D homogeneous GRF with the covariance

\begin{equation}\label{cov}
\mathcal{E}(x,y)\equiv \left\langle \phi (\mathbf{x}^{\prime })\phi (\mathbf{%
	x}^{\prime }+\mathbf{x})\right\rangle =\exp \left( -\frac{x^{2}}{2\lambda
	_{x}^{2}}-\frac{y^{2}}{2\lambda _{y}^{2}}\right) ,
\end{equation}%
which yields the associated Blob functions
\begin{equation}
F_{B}(x,y)=\sqrt{\frac{\pi \lambda _{x}\lambda _{y}}{2}}\exp \left( -\frac{%
	x^{2}}{\lambda _{x}^{2}}-\frac{y^{2}}{\lambda _{y}^{2}}\right) .
\label{blob}
\end{equation}%
The correlation lengths are chosen $\lambda _{x}=0.3,\lambda _{y}=0.4.$ The Fourier's compact support is a rectangle in which $|\mathbf{k}_{i}|\lambda _{i}\leq 5$ which ensures that $99.91\%$ of the spectrum is reproduced. The Blob's compact support is a rectangle in which $|\mathbf{a}_{i}|/\lambda _{i}\leq 4$ which ensure that $99.99\%$ of the covariance is reproduced. 

More precisely, we investigate numerically the effects of the additional stochastic elements introduced in the representations \eqref{eq_0.3a} and \eqref{eq_0.3b} and the ability of the simple discrete distribution \eqref{616} to improve the convergence rate. For that, we shall use, further, six methods of computing the GRF, which are described in Table \ref{table0}. The notation for each type of representation consists of three characters: the first letter for the method (Fourier or Blob), the second for how the parameters are distributed (Fixed or Random) and the third for the distribution of the random function $\zeta $ (Continuous or Discrete distributions).

\begin{table}
	\centering
	\begin{tabular}{|c|c|c|c|c|c|c|c|}
		\hline
		\text{} & \text{FFC} & \text{FRC} & \text{FRD} & \text{BFC} & \text{BRC} & 
		\text{BRD} \\ \hline
		\text{Method} & \text{Fourier} & \text{Fourier} & \text{Fourier} & \text{Blob%
		} & \text{Blob} & \text{Blob} \\ \hline
		\text{Parameters} & Fixed & Random & Random & Fixed & Random & Random \\ \hline
		$\zeta$ & $[0,8)$ & $[0,8)$ & $\pm 1$ & Gaussian & Gaussian & $\pm1$ \\ 
		\hline
	\end{tabular}
	\captionof{table}{The numerical representations tested in this Section.} %
	\label{table0}
\end{table}

\subsection{Reproducing the covariance}

\begin{figure}[tbp]
	\centering
	\subfloat{\includegraphics[width = 0.5\linewidth]{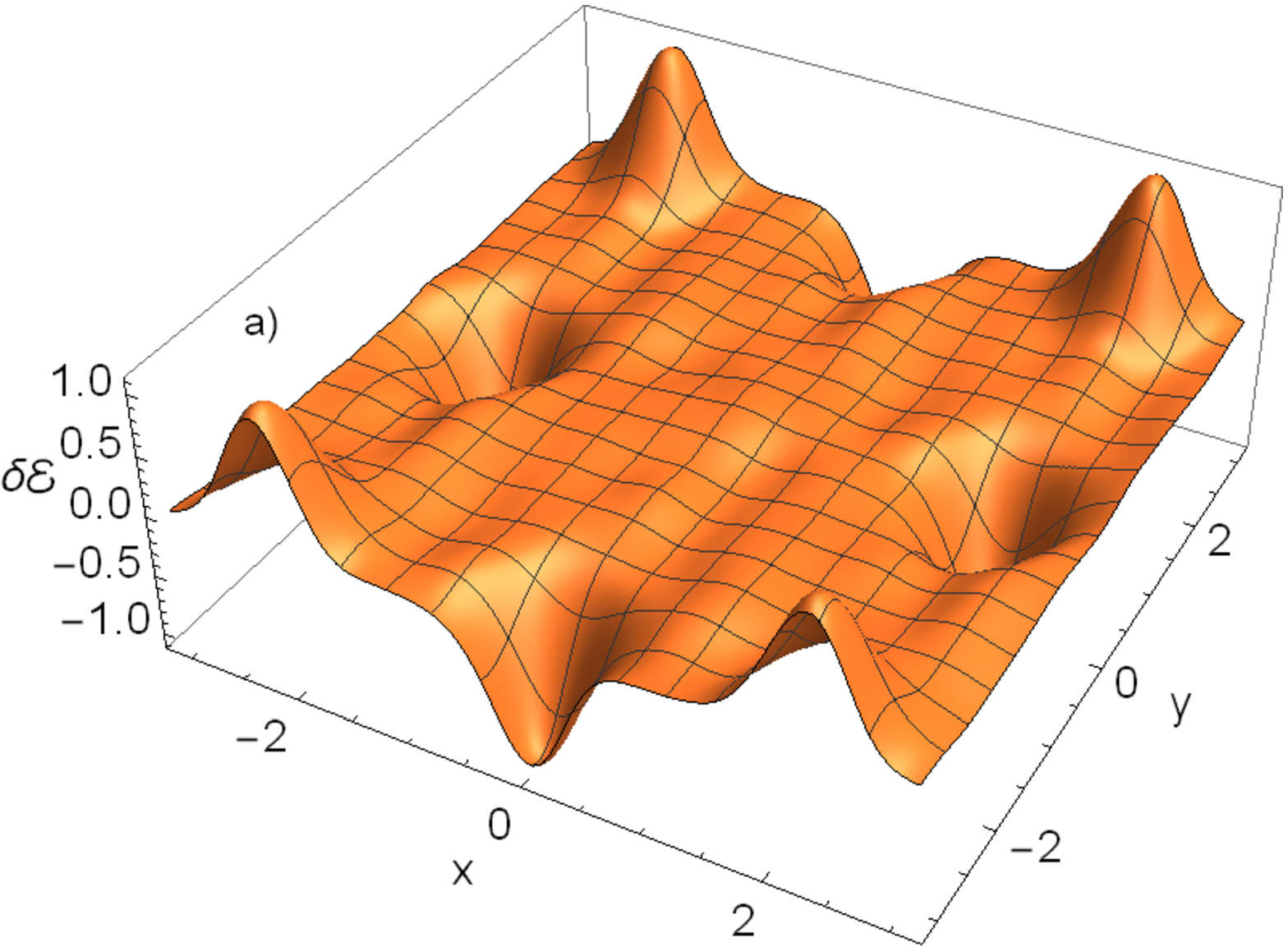}} %
	\subfloat{\includegraphics[width = 0.5\linewidth]{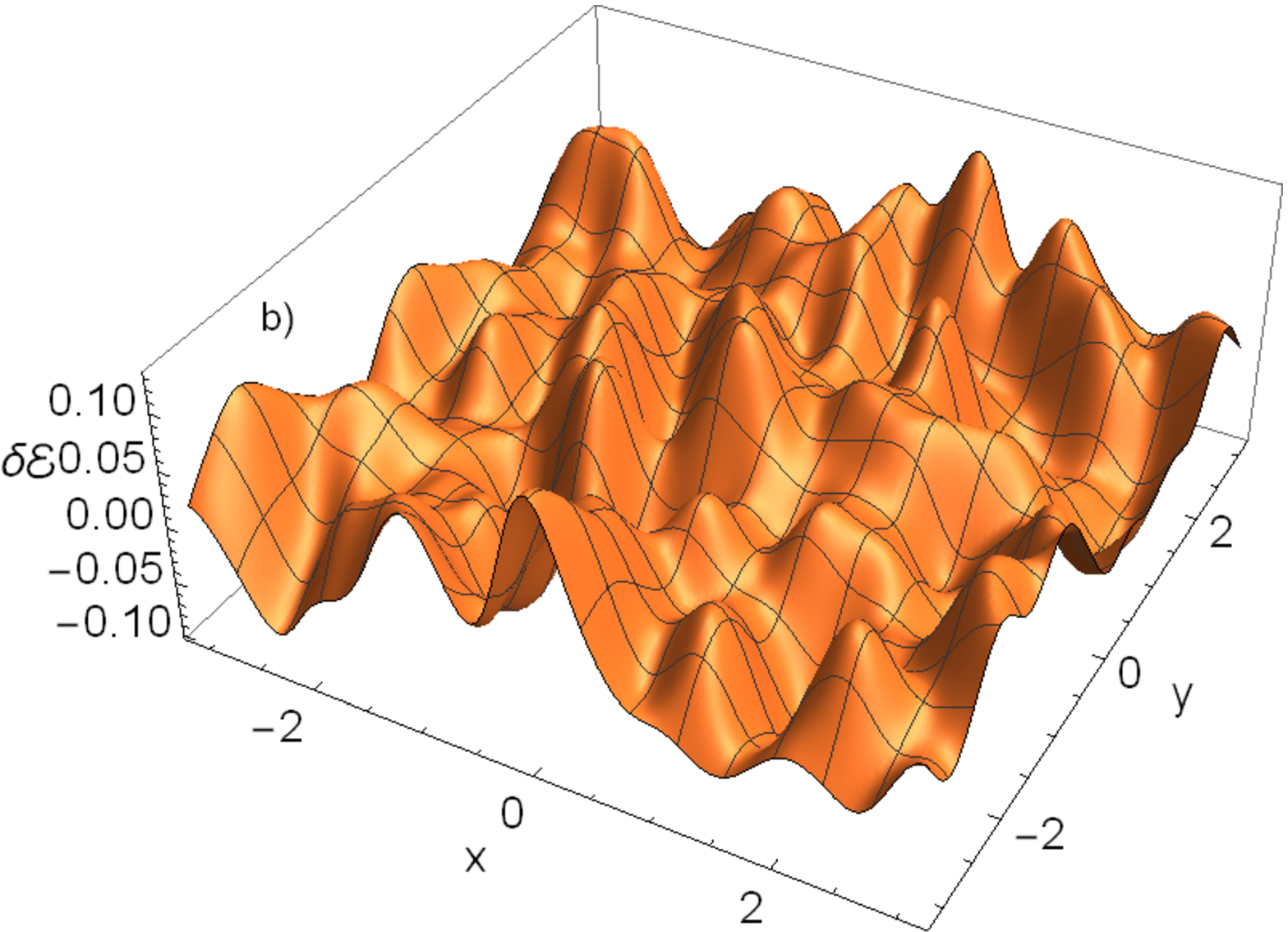}}\newline
	\subfloat{\includegraphics[width = 0.5\linewidth]{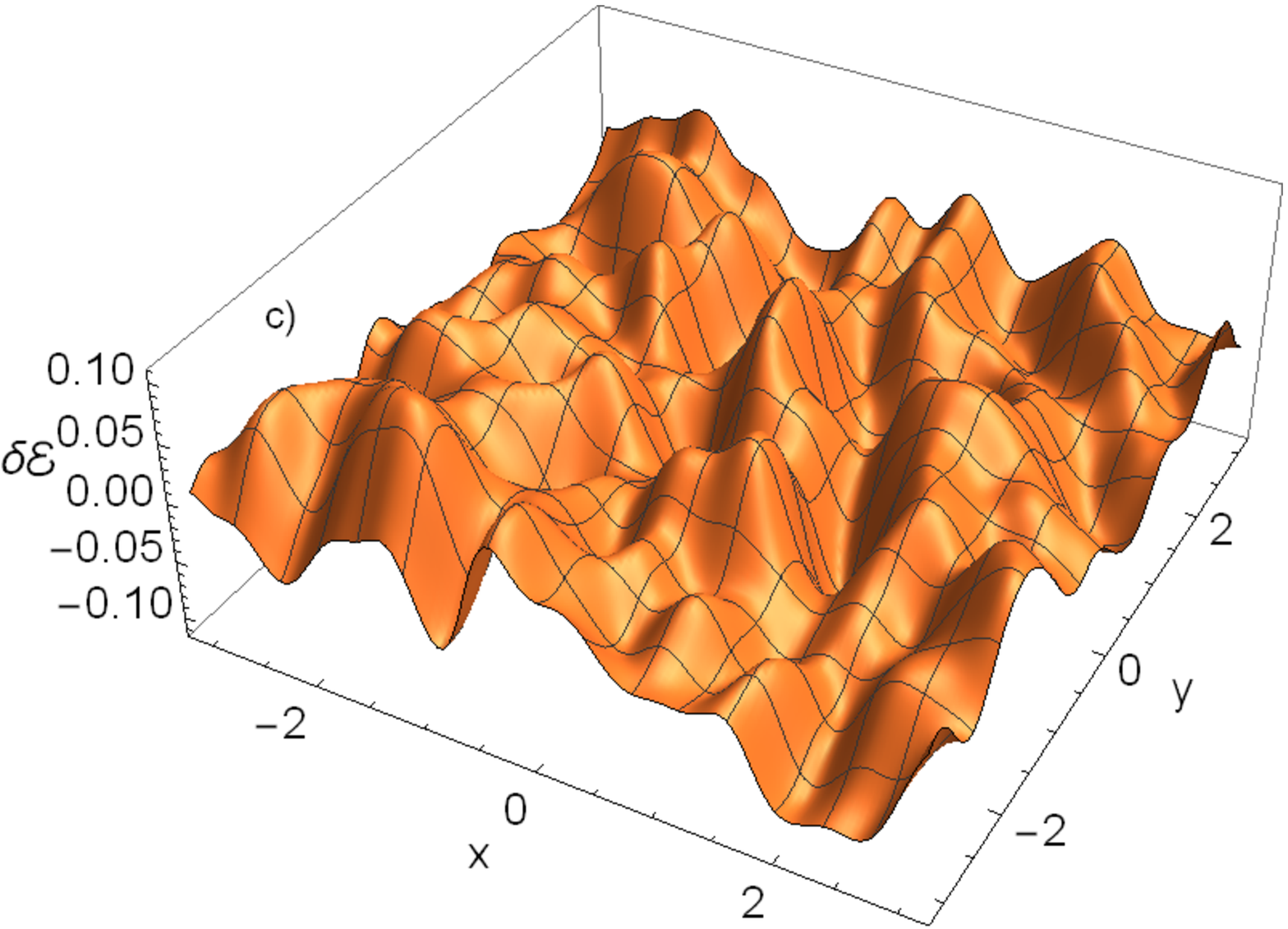}} %
	\subfloat{\includegraphics[width = 0.5\linewidth]{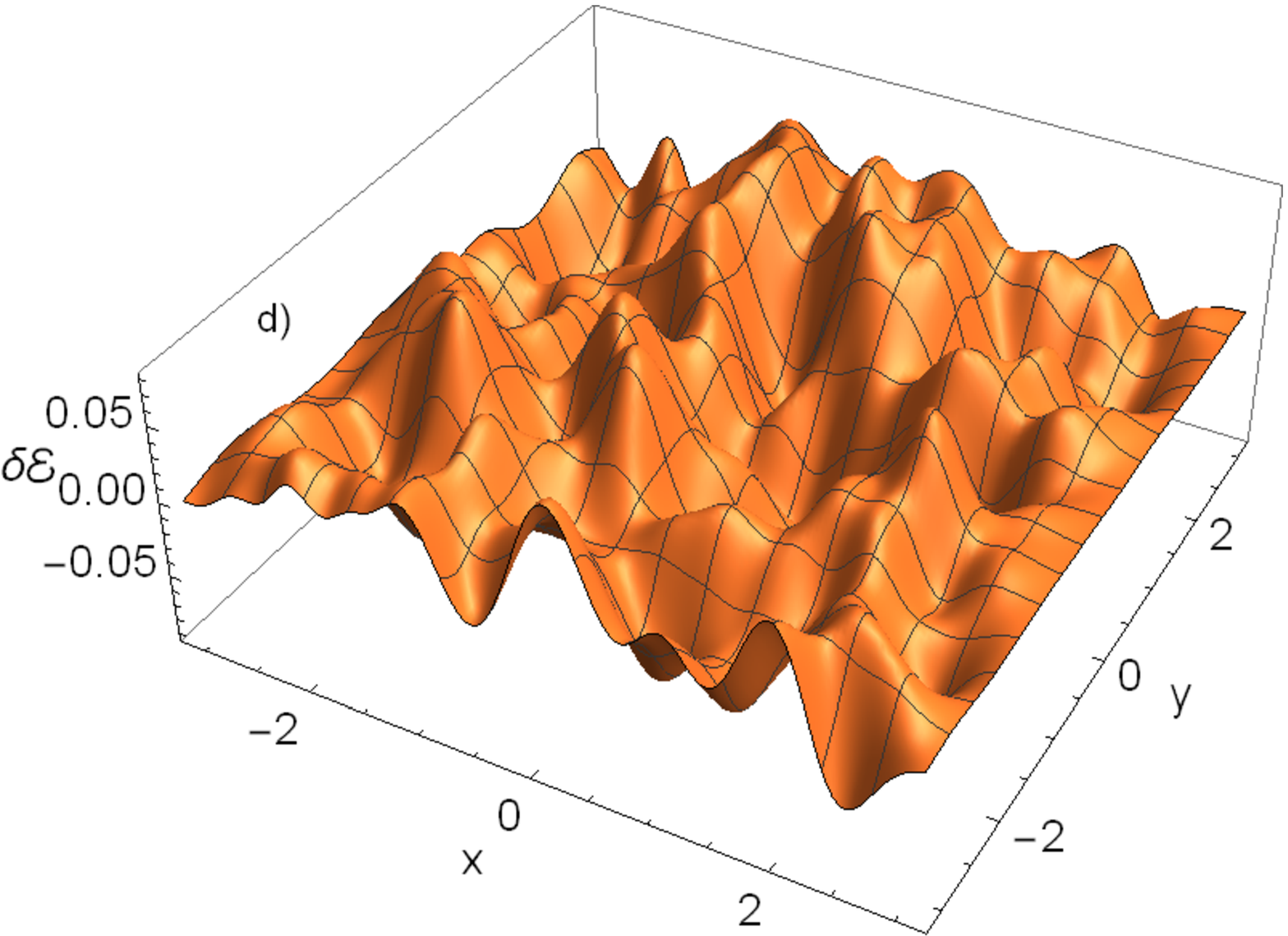}}\newline
	\subfloat{\includegraphics[width = 0.5\linewidth]{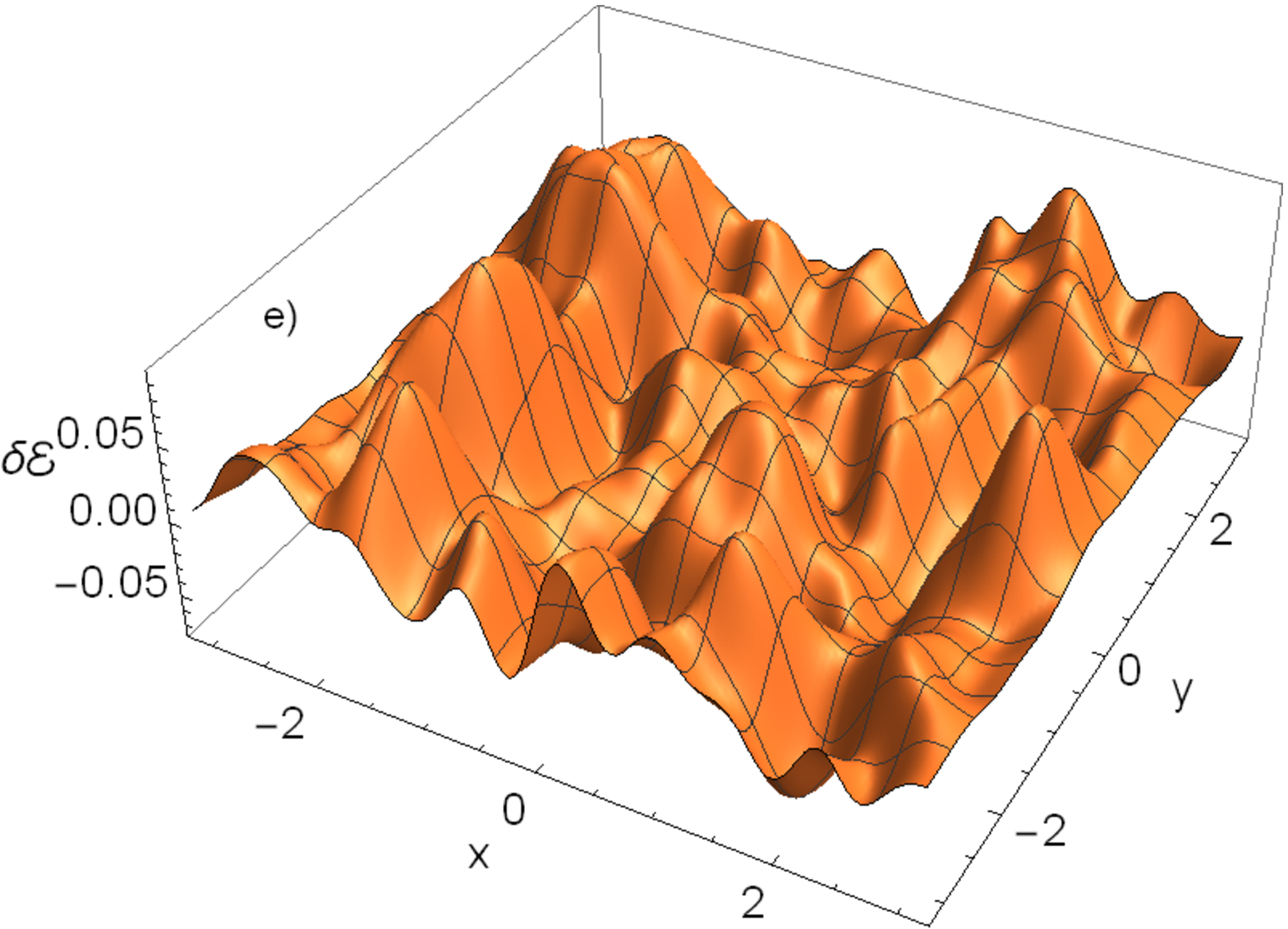}} %
	\subfloat{\includegraphics[width = 0.5\linewidth]{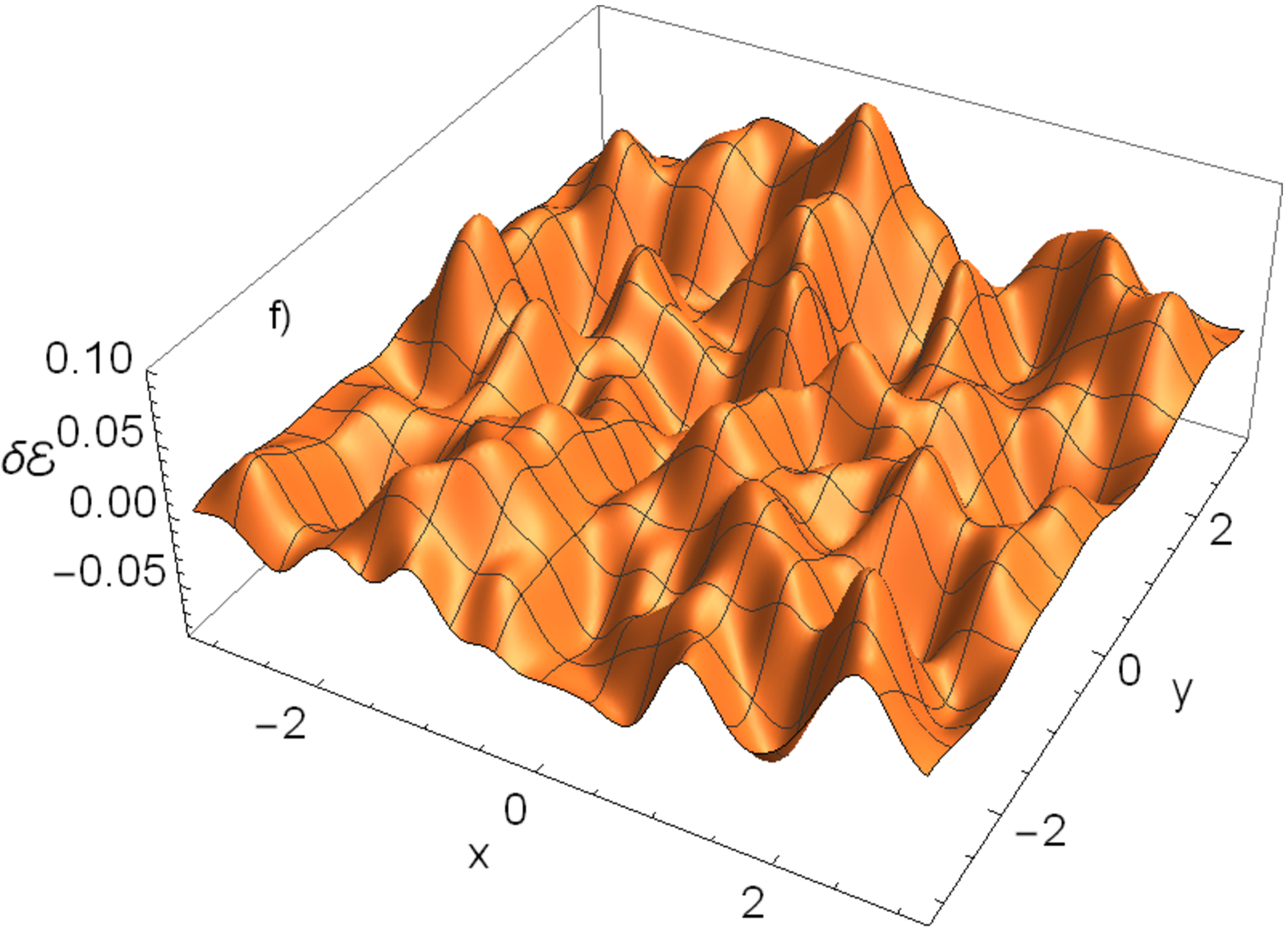}}\newline
	\caption{The error $\delta \mathcal{E}(\mathbf{x})$ of the covariances averaged over ensembles of $M=1000$
		realizations with the FFC (a), FRD(b), FRC (c), BFC (d), BRD(e), BRC (f)
		methods.}
	\label{Fig_2}
\end{figure}

We construct an ensemble of $M=10^{3}$ realizations of the GRF $\phi (\mathbf{x})$ with the covariance \eqref{cov} on a rectangular domain $[-\pi ,\pi]\times \lbrack -\pi ,\pi ]$ using all six methods. A small number of parametric points $N_c=12^{2}$ in the compact support was chosen for each method. The fluctuations of the resulting covariance around the exact profile $\delta \mathcal{E}(\mathbf{x})=\langle \phi (\mathbf{0})\phi (\mathbf{x})\rangle -\mathcal{E}(\mathbf{x})$ can be seen in Fig.\ref{Fig_2}. All methods offer similar amplitudes except the FFC method, which, due to its fixed equidistant grid in the $\mathbf{k}$ space has an unphysical periodicity.

\begin{figure}[tbp]
	\centering
	\subfloat{\includegraphics[width = 1.\linewidth]{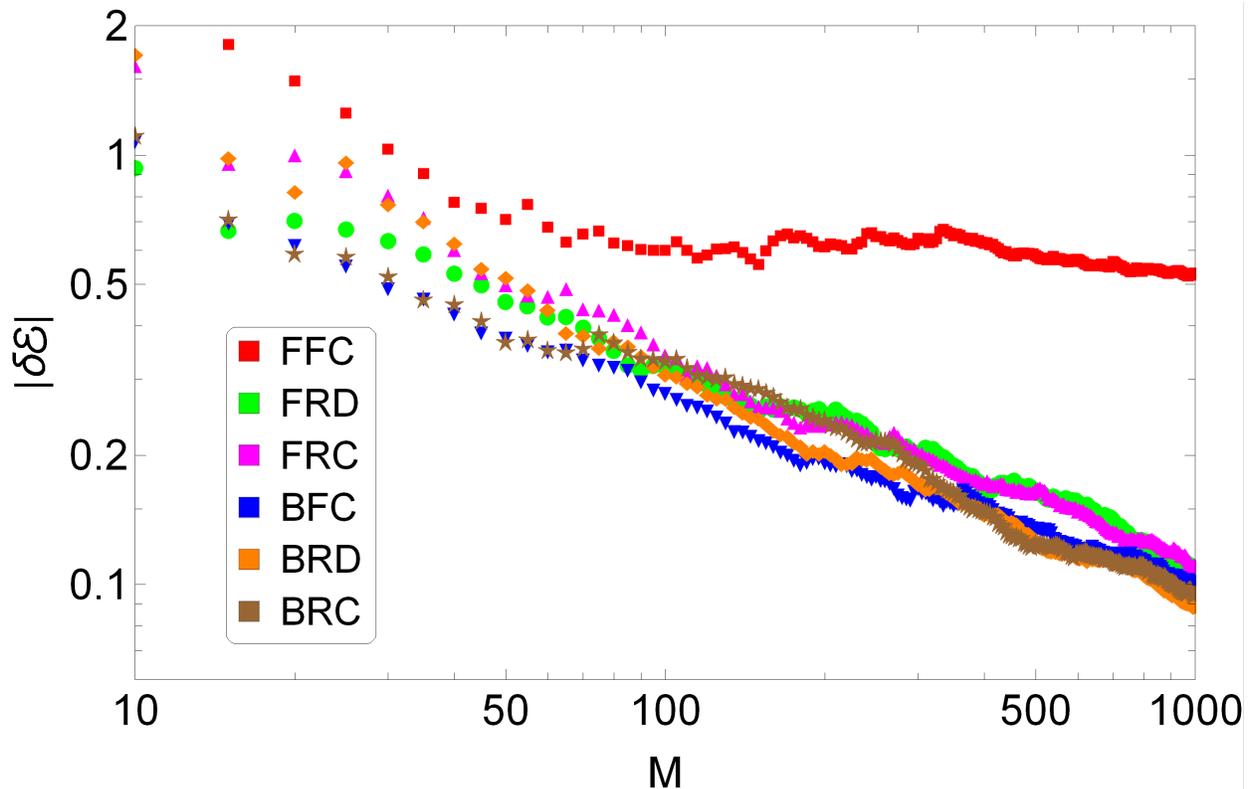}}
	\caption{Evolution of covariance error with the dimension of the ensemble for the FFC (red square), FRD (green circle), FRC (magenta up-triangle), BFC (blue down triangle), BRD (orange diamond), BRC (brown star) methods.}
	\label{Fig_3}
\end{figure}

The rate of convergence for the error of the covariance function $|\delta \mathcal{E}|=\int |\mathcal{E}(\mathbf{x})-\mathcal{E}_{approx}(\mathbf{x})|d\mathbf{x}$ can be seen in Fig. \ref{Fig_3} ($N_c=12^{2}$) as function of $M$, the ensemble dimension. One can see that $|\delta \mathcal{E}|$ decays with the increase of $M$ at approximately the same rate for five of the above methods and that the FFC (standard FFT) has a much weaker convergence at small values of $N_c.$ These five methods are able to reproduce the covariance even at small values of the number of elements in the sums in Eqs. \eqref{eq_0.3a},\eqref{eq_0.3b}. On the contrary, the FFC method (standard FFT) offers a poor representation of the covariance function on grids with low densities of points, in comparison with the other proposed methods. Increasing $N_c$, the decay rate of the error increases for the FFC method, but values similar to the other representation are attained at very large $N_c$ (of the order $\sim 100^2$). Essentially, the fail is due to the weak stochastic character of FFC (fixed grid for the wave numbers). We note that the corresponding fixed grid Blob method (the BFC) gives much better results in spite of the same weak stochastic character.

Thus, reasonable values of the error of the convolution are obtained with the FFC method at much larger values of M and/or $N_c$. The computational time that scales as $M\times N_c$ is much longer for the FFC than for the other five methods (by at least one order of magnitude).

\subsection{Reproducing the Gaussian character}

\begin{figure}[tbp]
	\centering
	\subfloat{\includegraphics[width = 0.99\linewidth]{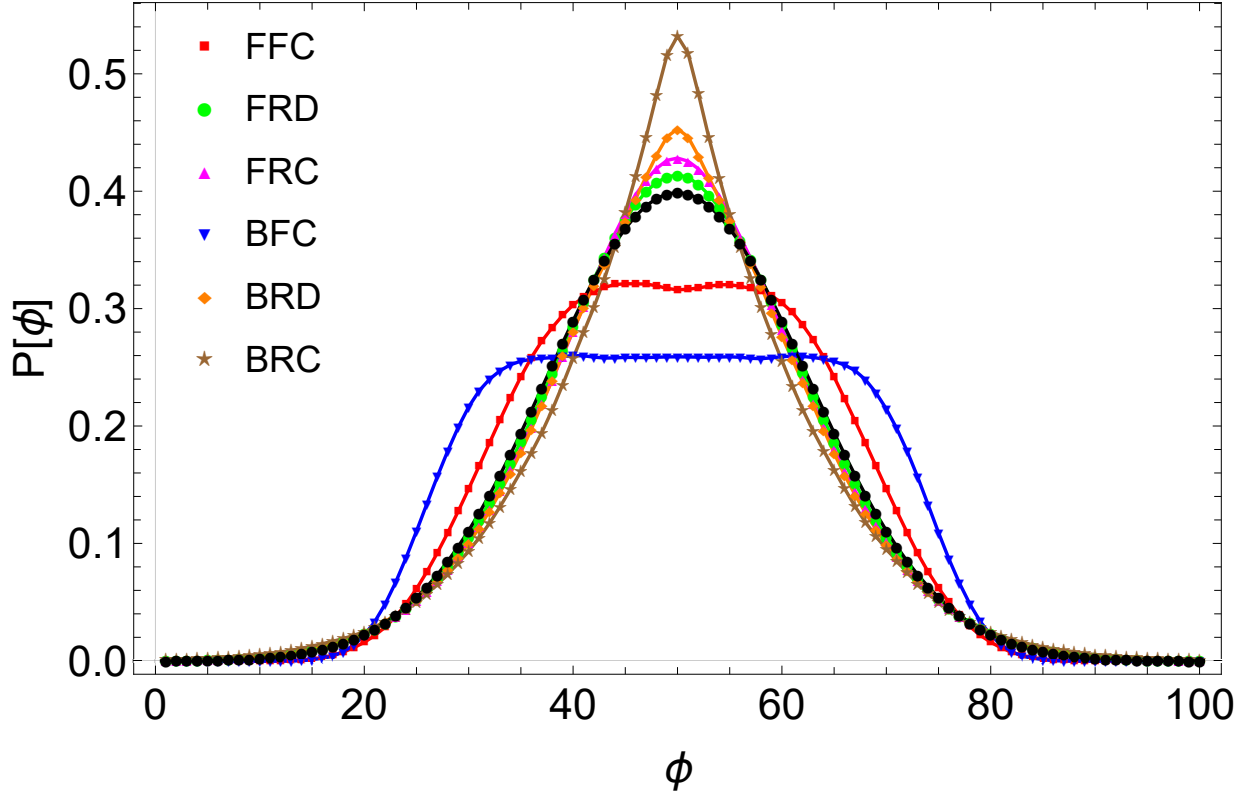}}
	\caption{PDFs of $\protect\phi(\mathbf{0})$ obtained with the FFC (red square), FRD (green circle), FRC (magenta up-triangle), BFC (blue down triangle), BRD (orange diamond), BRC (brown star) methods and a statistical ensemble of $M=10^7$ realizations.}
	\label{Fig_4b}
\end{figure}

We have generated large ensembles ($M=10^{7}$) of GRFs with all methods using even fewer points $N_c=5^{2}$. In order to test the Gaussianity of the resulting fields, we have focused mainly on the one-point PDF of the field $\phi (\mathbf{0})$. We note that much larger values of M are necessary in order to reduce the statitical fluctuations in the computed PDFs. The results are presented in Fig. \ref{Fig_4b}. One can see that the FFC has low quality for the potential distribution, as for the covariance function (Fig. \ref{Fig_2}). The corresponding Blob representation (with fixed grid, BFC) is even worse for $P(\phi)$ (see Fig. \ref{Fig_4b}). It is obvious that the use of random grids instead of fixed ones is a much better choice also in the matter of Gaussianity. Moreover, as it has been stated in Section \ref{section_1.2}, using discrete distributions $\zeta _{j}=\pm 1$, instead of distributions with continuous support, offers significant improvements in the profile $P(\phi )$: FRD and BRD are better than FRC and BRC.

Table \ref{table1} quantifies these results computing the first even moments for the PDF of $\phi (\mathbf{0})$. The global error defined as $|\delta P|=\int |P_{gauss}(\phi)-P_{method}(\phi )|d\phi $ and the 4-point correlation function $\mathcal{E}_{(4)}=\langle \phi (\mathbf{x}_{1})\phi (\mathbf{x}_{2})\phi (\mathbf{x}_{3})\phi (\mathbf{x}_{4})\rangle $ with $\mathbf{x}_{1}=(0,0),$ $\mathbf{x}_{2}=(\lambda _{x},\lambda _{y}),$ $\mathbf{x}_{3}=(\lambda _{x}/2,-\lambda_{y}/3),$ $\mathbf{x}_{4}=(-\lambda _{x}/3,\lambda _{y}/2)$ \ are also shown. One can see that fixed grids lead to sub-Gaussian distributions while random grids to over-Gaussian ones (longer tails).

\begin{table}
	\centering
	\begin{tabular}{|c|c|c|c|c|c|c|c|}
		\hline
		\text{} & \text{FFC} & \text{FRC} & \text{FRD} & \text{BFC} & \text{BRC} & 
		\text{BR1} & \text{exact} \\ \hline
		$<\phi^2>$ & 1.046 & 1.002 & 0.999 & 1.264 & 1.001 & 1.001 & 1 \\ \hline
		$<\phi^4>$ & 2.504 & 3.424 & 3.223 & 3.257 & 3.976 & 3.543 & 3 \\ \hline
		$<\phi^6>$ & 8.447 & 21.32 & 18.38 & 11.04 & 30.83 & 23.37 & 15 \\ \hline
		$<\phi^8>$ & 35.13 & 199.82 & 145.48 & 44.19 & 372.89 & 235.67 & 105 \\ 
		\hline
		$|\delta P|$ & 0.167 & 0.061 & 0.026 & 0.309 & 0.136 & 0.066 & 0 \\ \hline
		$\mathcal{E}_{(4)}$ & 0.744 & 1.041 & 0.948 & 0.694 & 0.906 & 0.850 & 0.789
		\\ \hline
	\end{tabular}
	\captionof{table}{Numerical values of the first odd moments of the PDF $P(\varphi;\mathbf{0})$ for all six methods considered.} \label{table1}
\end{table}

Thus, we have shown that the best choices for the representation of homogeneous GRFs are based on random grids with $\zeta =\pm 1$, i.e. on FRD \eqref{eq_0.3a} or BRD \eqref{eq_0.3b} methods. Further, by Blob representation we shall refer to BRD while by Fourier to FRD methods in the remaining part of this paper. Note that the Fourier is slightly better than the Blob method.

\subsection{DNS of stochastic transport}

We have proven until now that the Fourier \eqref{eq_0.3a} and Blob \eqref{eq_0.3b} representations offer the best convergence rates from the perspective of their Eulerian properties. Now, we perform additional tests regarding their Lagrangian abilities in the context of a DNS of a V-Langevin equation. The following model has been chosen:

\begin{equation}
\frac{d\mathbf{x}(t)}{dt}=\mathbf{v}(\mathbf{x}(t))=\hat{e}_{z}\times \nabla \phi (\mathbf{x}(t))+V_{d}\hat{e}_{y},  \label{eq_1.5}
\end{equation}
where $\phi (\mathbf{x})$ is a GRF and $V_{d}\hat{e}_{y}$ is an average velocity. This stochastic equation describes the dynamics of test particles under electrostatic turbulence in magnetically confined plasmas \cite{PhysRevE.63.066304,PhysRevE.54.1857} or for tracer transport in incompressible turbulent fluids. The stochastic potential is considered frozen, i.e. the covariance is time independent. The covariance function is \eqref{cov} with $\lambda _{x}=1,\lambda _{y}=2$.

We have chosen this transport model because the ensemble of solutions exhibits two invariants: a "local" one characteristic to each trajectory and a global one, characteristic to the entire ensemble. Both are a consequence of the null divergence $\nabla \cdot \mathbf{v}(\mathbf{x})=0$ property and of the homogeneity of the stochastic field. The equation of motion \eqref{eq_1.5} is of Hamiltonian type, with $\phi_{t}(\mathbf{x})=\phi (\mathbf{x})+V_{d}x$ the Hamiltonian function. The latter is invariant in each realization of the potential $\phi (\mathbf{x}),$ since the trajectories obtained from Eq. \eqref{eq_1.5} evolve on the contour lines of $\phi _{t}(\mathbf{x}).$\ At $V_{d}=0,$ $\mathbf{x}(t)$\ are closed and have periodic dependence on time, while at $V_{d}\neq 0$ some of the trajectories are opened. 

The second invariant is statistical and involves the Lagrangian velocity $\mathbf{v}(\mathbf{x}(t)).$ According to Lumley's Theorem \cite{monin1971statistical,PhysRevE.66.038301}, the statistics of the Lagrangian
velocity is identical with the statistics of the Eulerian velocity, at any time 
\begin{equation*}
P^{L}[\mathbf{v}(\mathbf{x}(t))]=P^{E}[\mathbf{v}(\mathbf{x})],
\end{equation*}%
where $P^{L}=\left\langle \delta[ \mathbf{v}-\mathbf{v}(\mathbf{x}(t))]\right\rangle $ is the Lagrangian probability and $P^{E}=\left\langle\delta[ \mathbf{v}-\mathbf{v}(%
\mathbf{x})]\right\rangle $ is the Eulerian probability. The latter is a space-independent Gaussian function 
\begin{equation*}
P^{E}(\mathbf{v})=\exp \left( -\frac{v_{x}^{2}}{2V_{xx}}-\frac{%
	(v_{y}-V_{d})^{2}}{2V_{yy}}\right) ,
\end{equation*}%
where $V_{ii}=\langle v_{i}(\mathbf{0})v_{i}(\mathbf{0})\rangle =-\partial_{jj}\mathcal{E}(\mathbf{x})|_{\mathbf{x}\rightarrow \mathbf{0}}=1/\lambda_{j}^{2}$.

The existence of the constraints related to these invariants makes the transport process very complicated, but it also provides strong benchmarks for the numerical simulations.

Regarding the numerical implementation, a second order Runge-Kutta numerical integration scheme has been used for a time interval of $[0,t_{max}]=[0,40]$ with a fixed time step $dt=0.04$. An ensemble of $M=3\times 10^{4}$ trajectories has been resolved. We have implemented the Fourier (FRD) and Blob (BRD) representations with $N_c=6^2,12^2$. Two cases have been considered: $V_{d}=0$ and $V_{d}=0.4$. A Fourier simulation with $N_c$ waves is denoted by $FN_c$ while a Blob one by $BN_c$ where $N_c=36$ or $144.$ 

We underline that $M,$ the dimension of the ensemble and (especially) $N_c,$ the number of random parameters in the series \eqref{eq_0.3a}, \eqref{eq_0.3b}, are small compared to the usual values in DNS. Thus, the DNS can be performed on personal computers, where the typical running times are rather small, of the order of $t_{CPU}\sim 10^{2}s$. 

\begin{figure}[tbp]
	\centering	\label{fig:traja} 
	\includegraphics[width =0.75\linewidth]{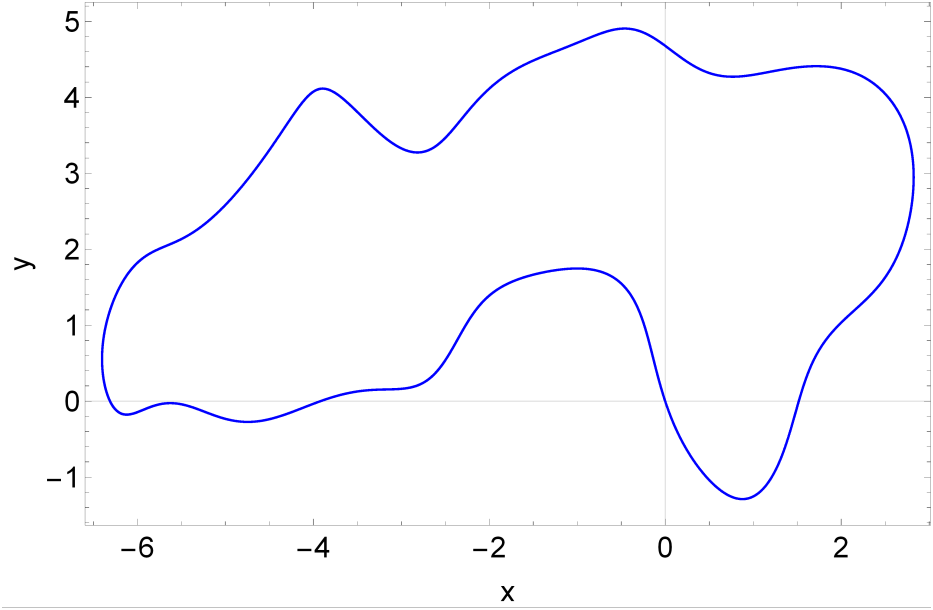}
	\caption{
		A typical trajectory simulated over $10$ periods.
	}
	\label{fig:trajb}
\end{figure}

First, we have checked that the numerical integration and the use of the generators \eqref{eq_0.3a},\eqref{eq_0.3b} of the GRF do not affect the Hamiltonian character of the trajectories. We plot in Fig. \ref{fig:trajb} a randomly chosen trajectory for almost $10$ times its period. Qualitatively, the trajectory remains closed. Apart from small oscillations $\delta \phi (t)/\bar{\phi}\sim 10^{-4}$, the potential is perfectly conserved along the represented trajectory. 
Thus, the combined errors from the approximation of the field and from the numerical integration remain small.

\begin{figure}[tbp]
	\centering
	\subfloat{\includegraphics[width = 0.45\linewidth]{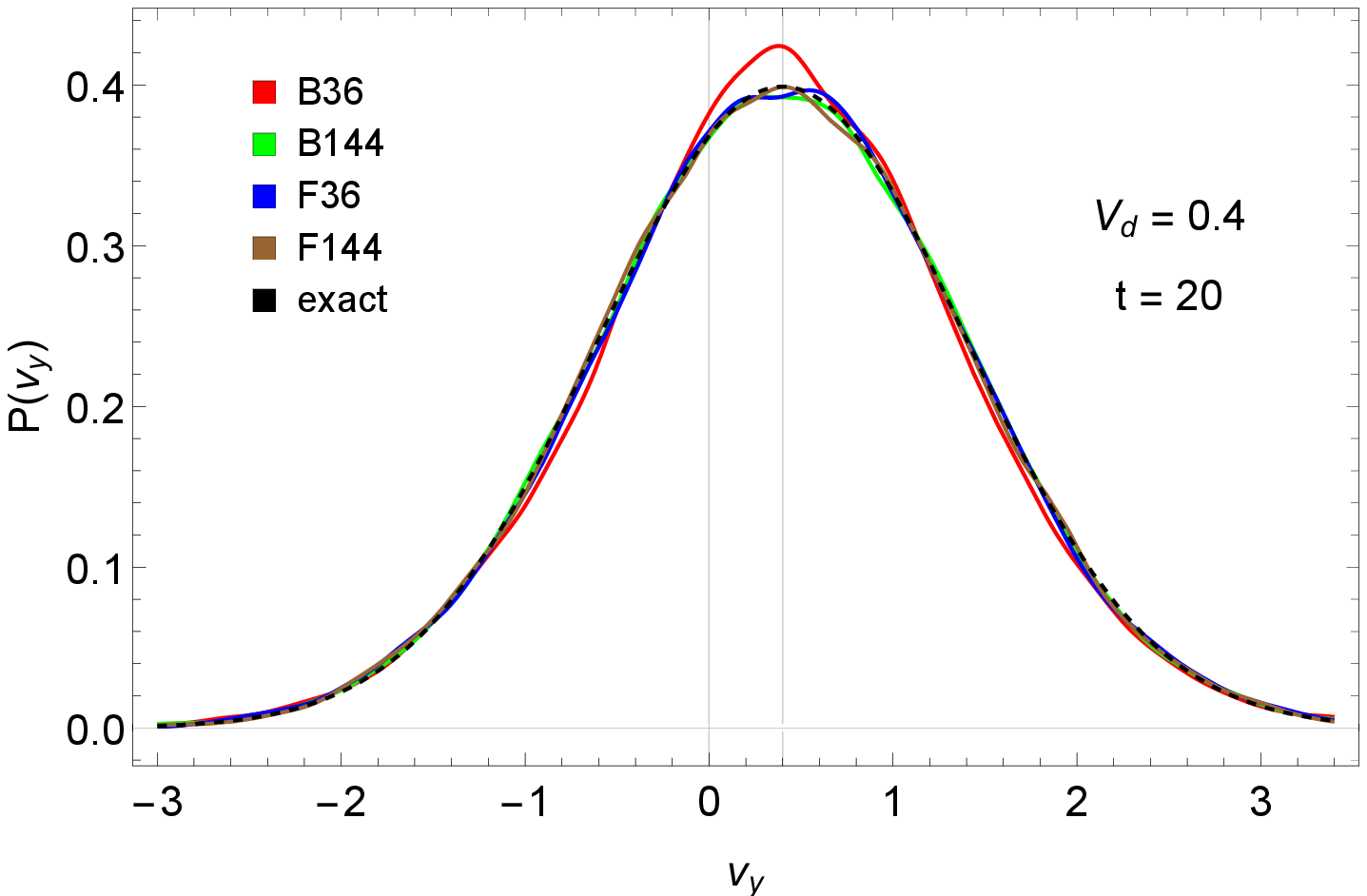}}
	\subfloat{\includegraphics[width = 0.45\linewidth]{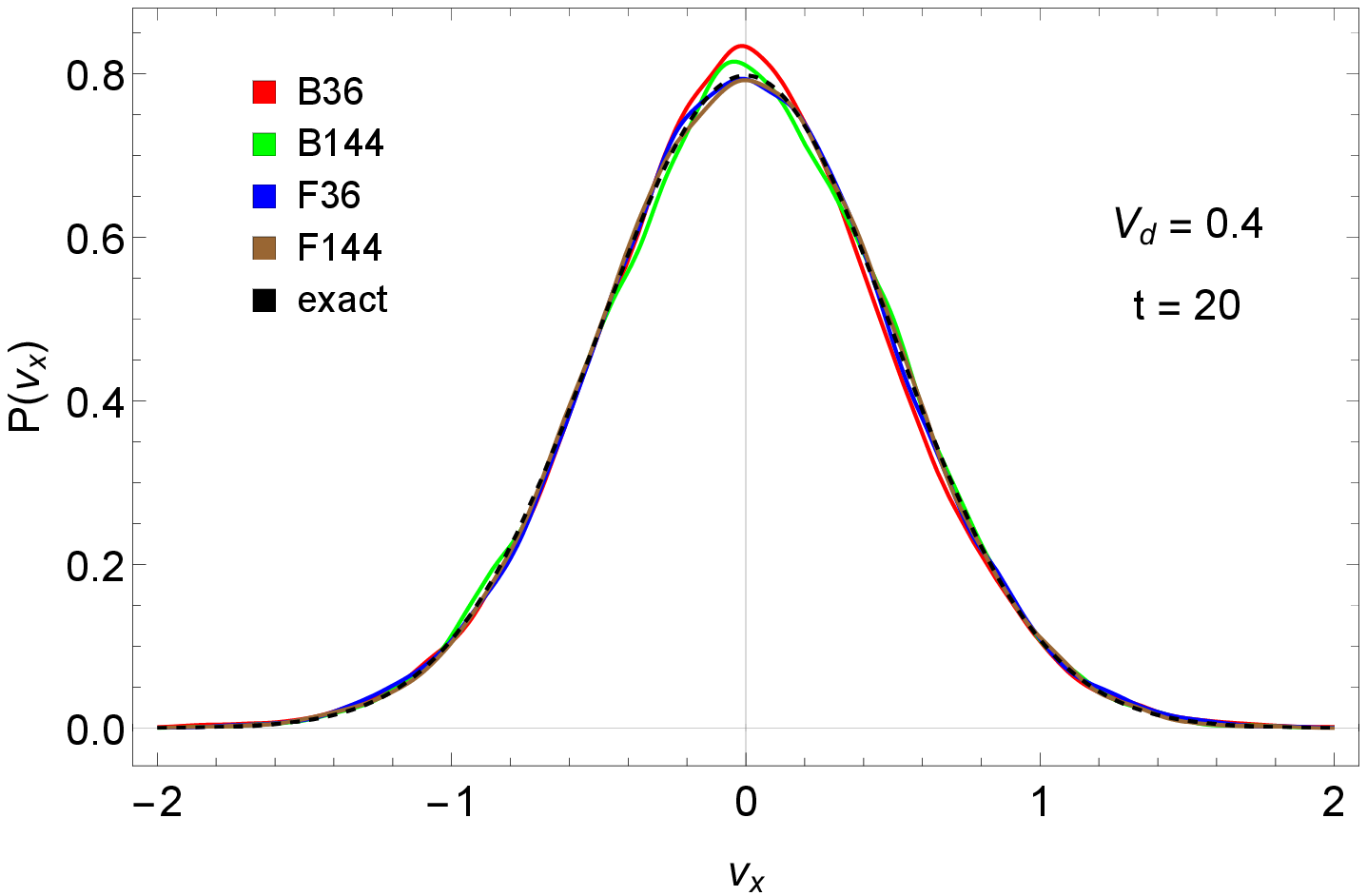}}
	\caption{Large time $t=20$ distribution of velocities. The exact, Gaussian
		shapes are in dashed lines.}
	\label{fig:32}
\end{figure}

\begin{figure}[tbp]
	\centering
	\subfloat{\includegraphics[width = 0.45\linewidth]{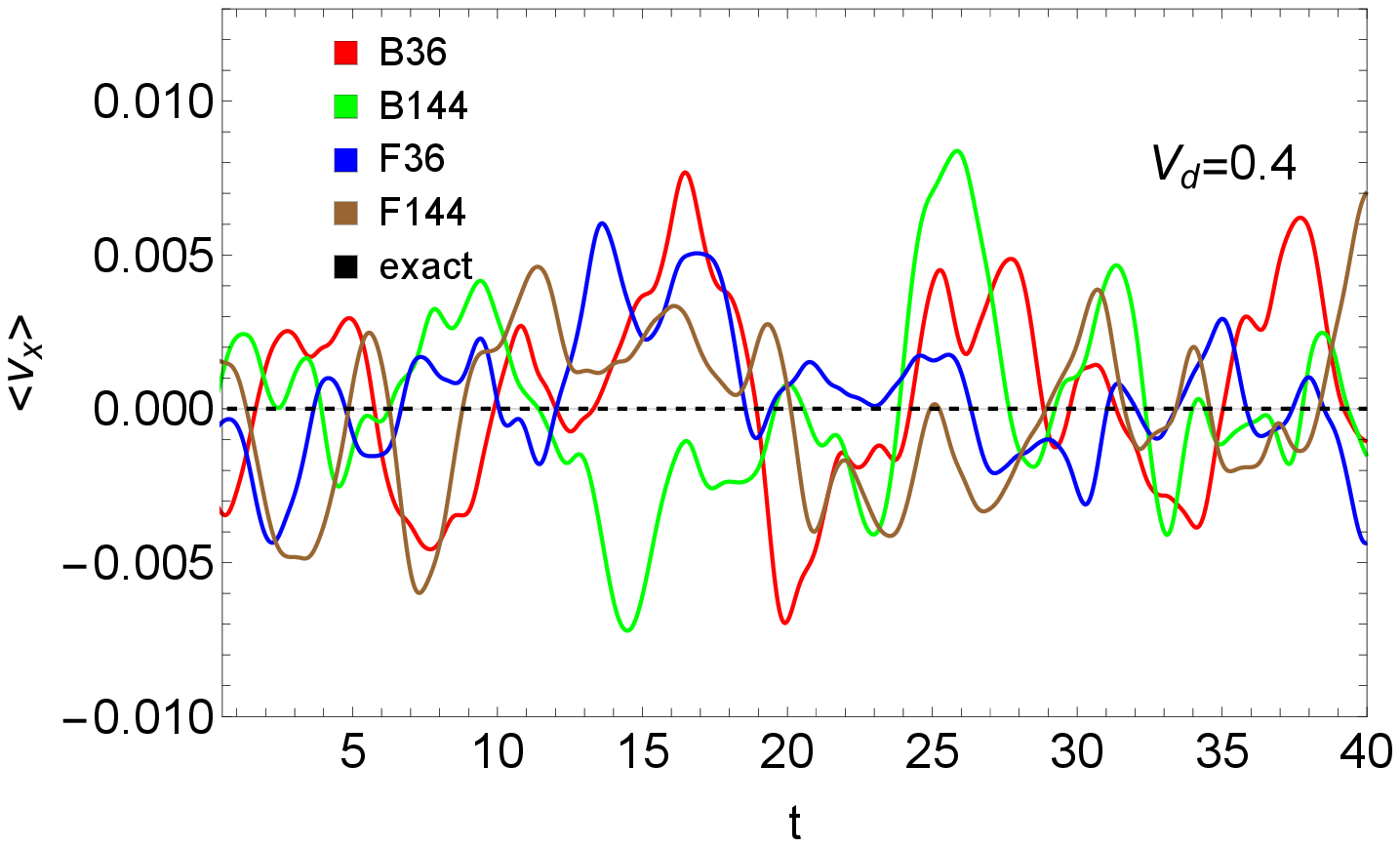}} %
	\subfloat{\includegraphics[width = 0.45\linewidth]{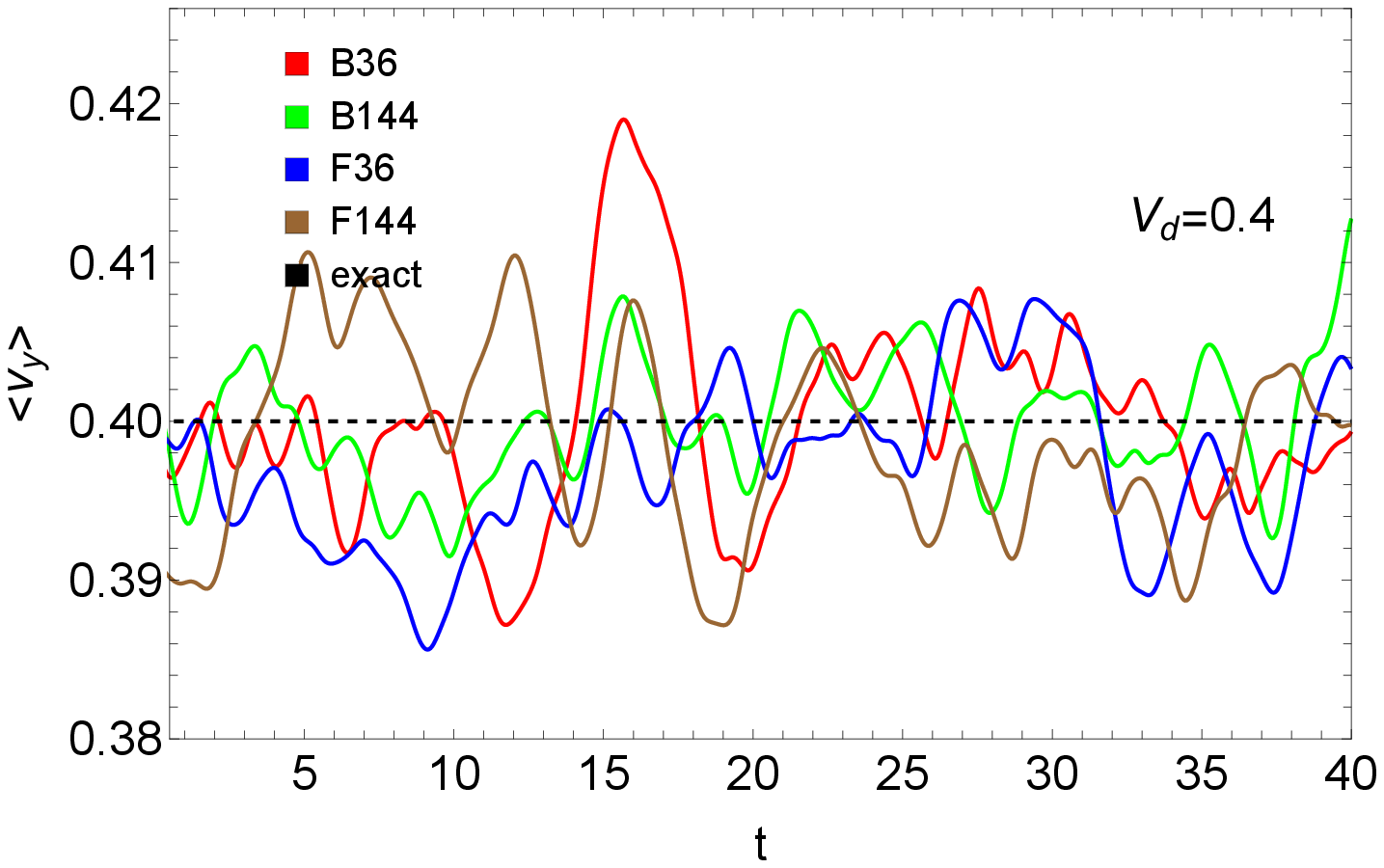}}\newline
	\subfloat{\includegraphics[width = 0.45\linewidth]{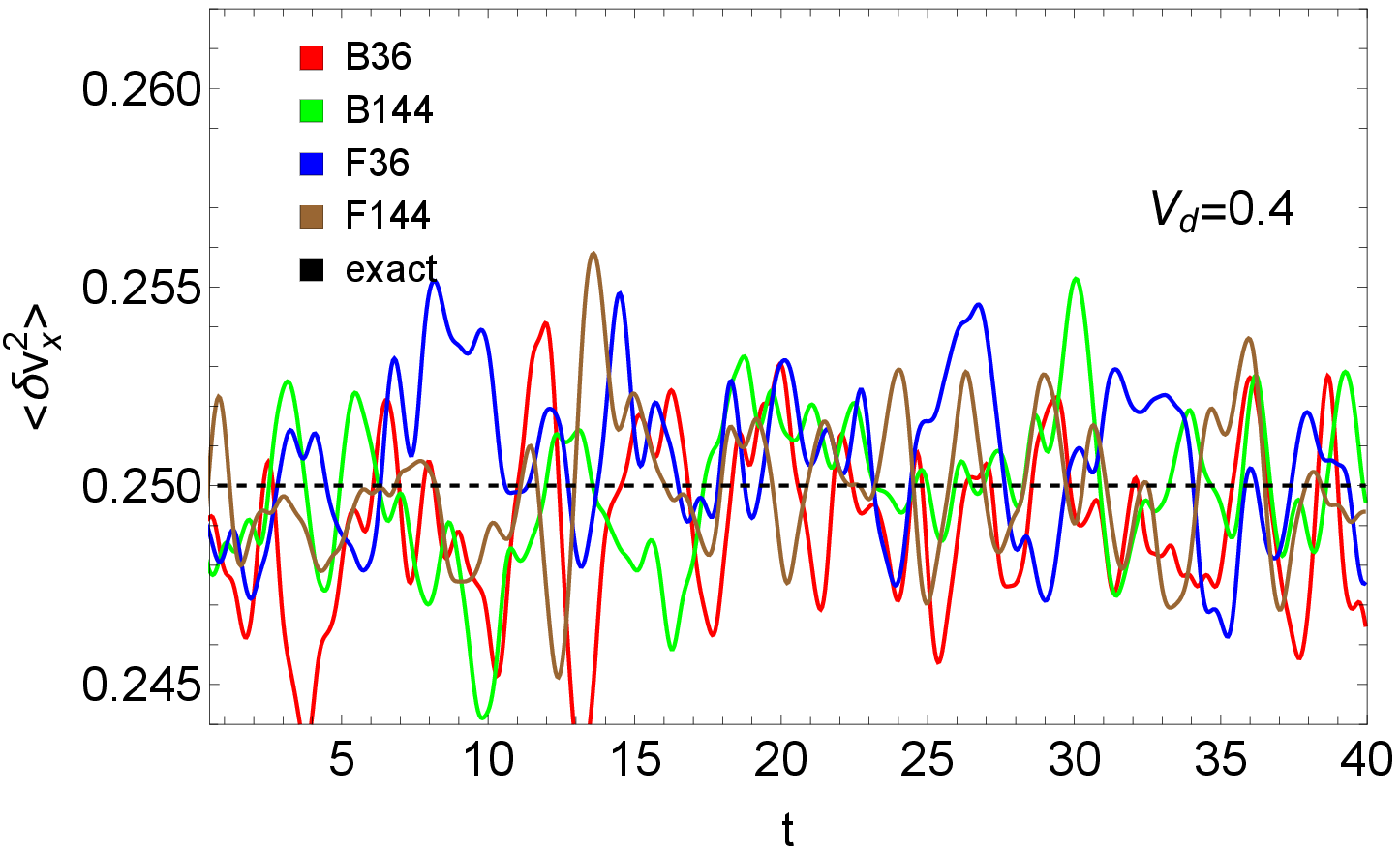}} %
	\subfloat{\includegraphics[width = 0.45\linewidth]{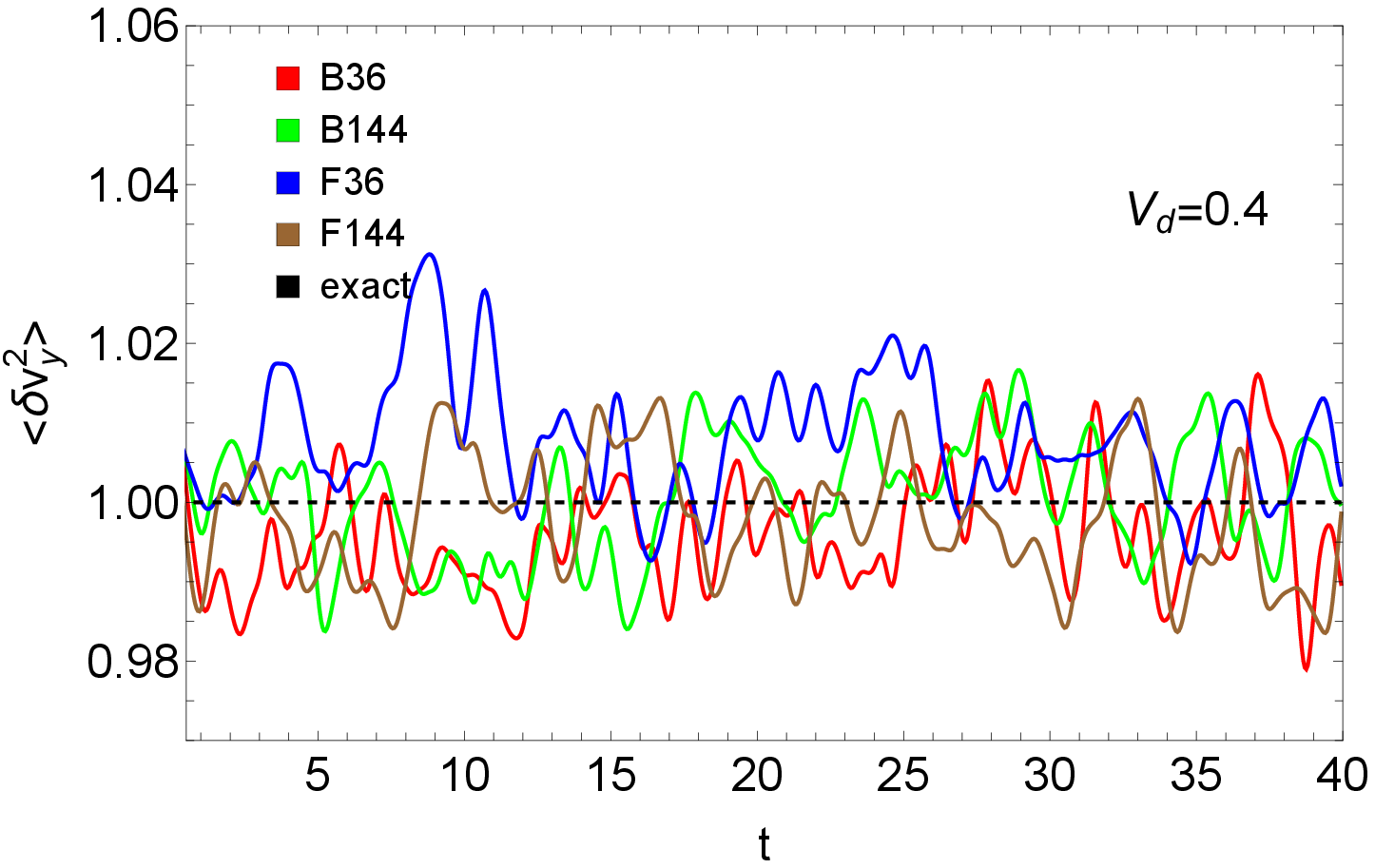}}\newline
	\subfloat{\includegraphics[width = 0.45\linewidth]{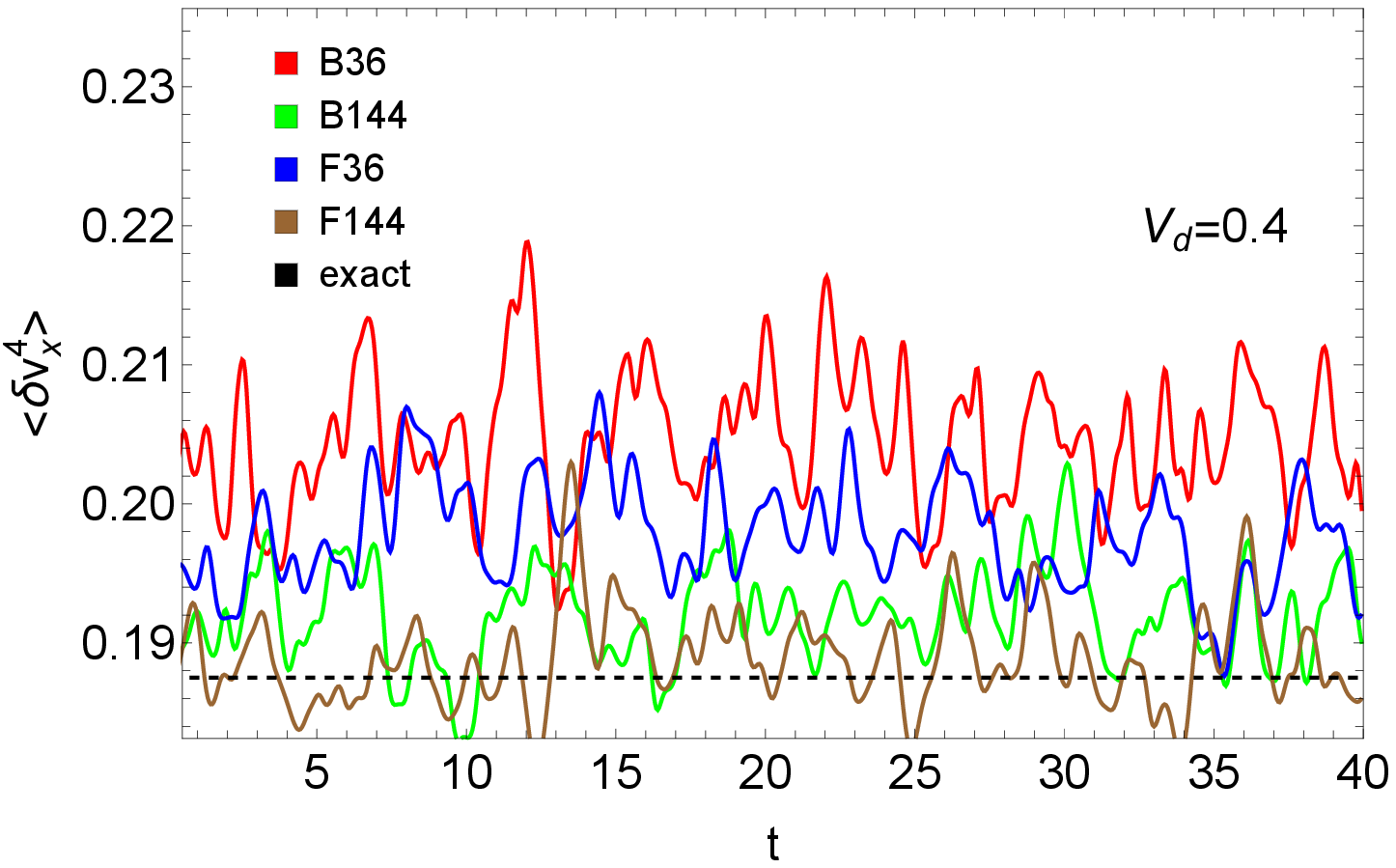}} %
	\subfloat{\includegraphics[width = 0.45\linewidth]{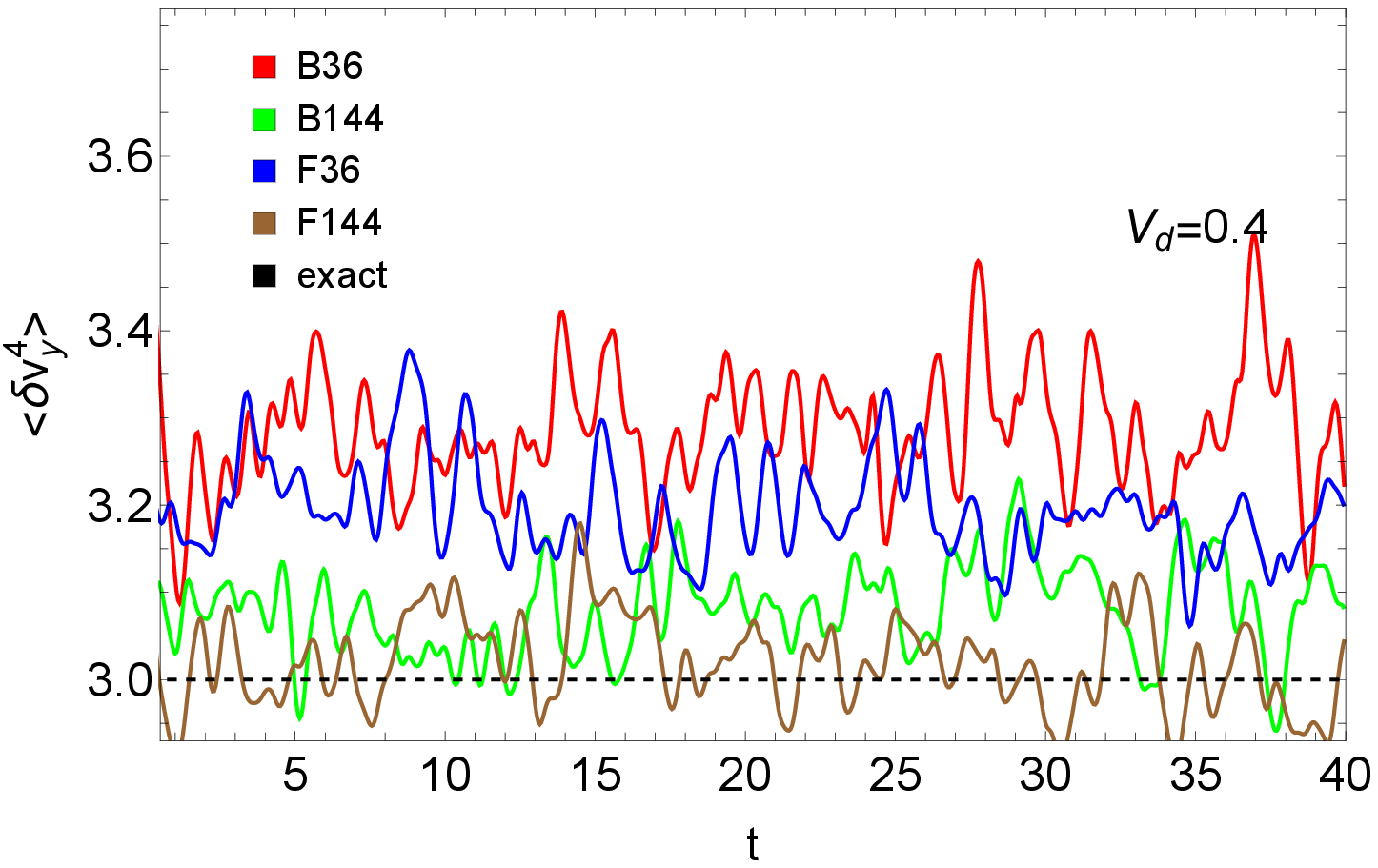}}
	\caption{Time dependence of average Lagrangian velocity moments.}
	\label{fig:33}
\end{figure}

Second, we test the global invariants by computing the PDF of Lagrangian velocities $P^{L}[\mathbf{v}(\mathbf{x}(t))]$ as well as its first moments $\langle v_{i}^{j}(t)\rangle ,~j=1,4$. Figure \ref{fig:32} shows the components of this distribution of the Lagrangian velocity at the moment $t=20,$ compared with the exact, Gaussian profiles. The results are close to the theoretical distributions, even for the cases $N_c=36.$ The statistical fluctuations can be analyzed more clearly in Figure \ref{fig:33} where  the moments $\langle v_{i}^{j}(t)\rangle $ are shown. On average, the Lagrangian invariance is well reproduced by both methods, the fluctuations being a consequence of a finite ensemble ($M=3\times 10^{4}).$ The deviations of the average values (for example $\langle \bar{v_{y}^{4}}\rangle \approx 3.3$ instead of the exact value $3$ for $B36$ method) are a consequence of a finite $N_c.$ As seen in Figure 6, the increase of $N_c$ approaches the averages to the theoretical values and reduces the statistical fluctuations. The results are satisfactory even at the small values taken here. 

\begin{figure}[tbp]
	\centering
	\subfloat{\includegraphics[width = 0.45\linewidth]{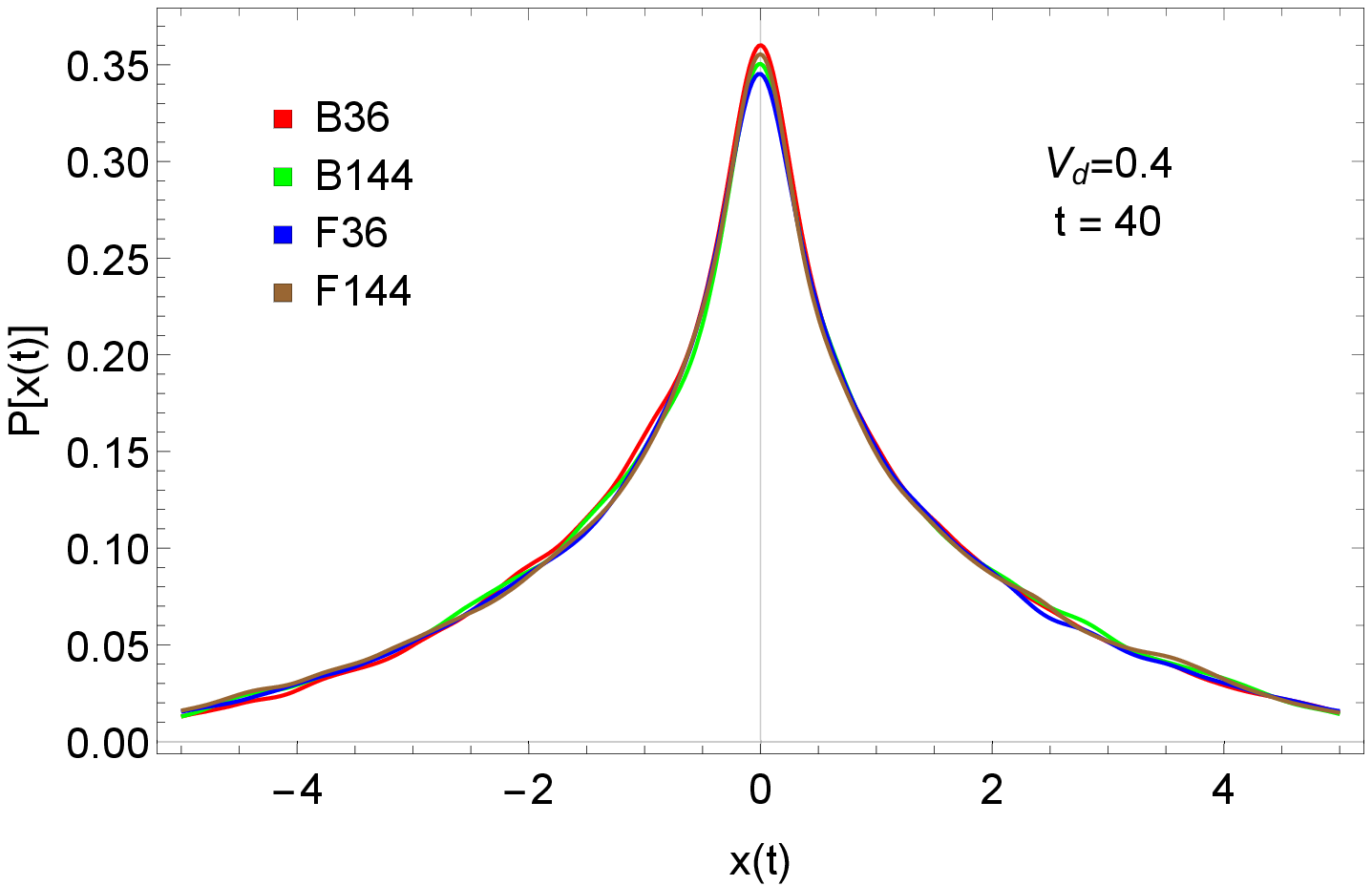}}
	\subfloat{\includegraphics[width = 0.45\linewidth]{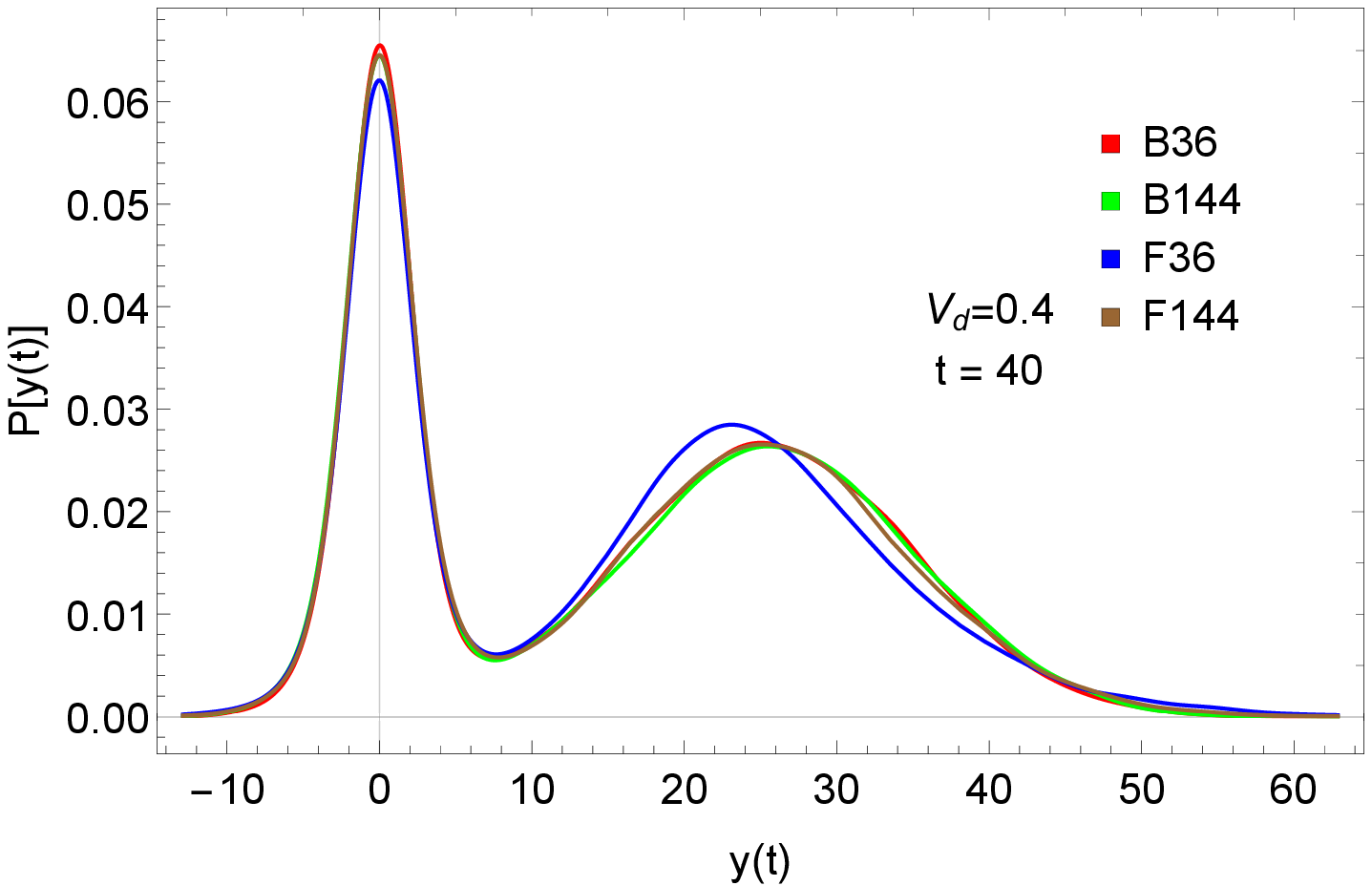}}
	\caption{Projected trajectory distributions $P(x(t)),P(y(t))$}
	\label{fig:21}
\end{figure}

A property of the stochastic transport described by Eq. \eqref{eq_1.5} is that the Gaussian, time-invariant Lagrangian velocity does not yield a Gaussian distribution of the trajectories $P(\mathbf{x}(t))=\langle \delta[\mathbf{x}-\mathbf{x}(t)]\rangle.$ The latter is a peaked function with long tails \cite{PhysRevE.70.056304}. This happens because of the trajectory trapping or eddying, produced by the invariance of the Lagrangian potential that ties particle paths on its contour lines \cite{PhysRevE.58.7359,PhysRevE.63.066304}. An average velocity opens a part of trajectories along its direction ($\hat{e}_{y}),$ but trapped particles still exist \cite{Vlad_2017}, as seen in Fig. \ref{fig:21} for $V_{d}=0.4.$ 

There are no clear theoretical results on $P(\mathbf{x(t)})$ that could be used as benchmark of the present DNS. Instead, we compare the results of the four runs commented here. 
The probability of the displacements $x(t)$ and $y(t)$ is shown in Fig \ref{fig:21}. The only observable difference appears in the distribution $P[y(t)]$ obtained in the $F36$ run. It underestimates the average displacement, the spreading of the free trajectories as well as the number of trapped particles.


\begin{figure}[tbp]
	\centering
	\subfloat{\includegraphics[width = 0.5\linewidth]{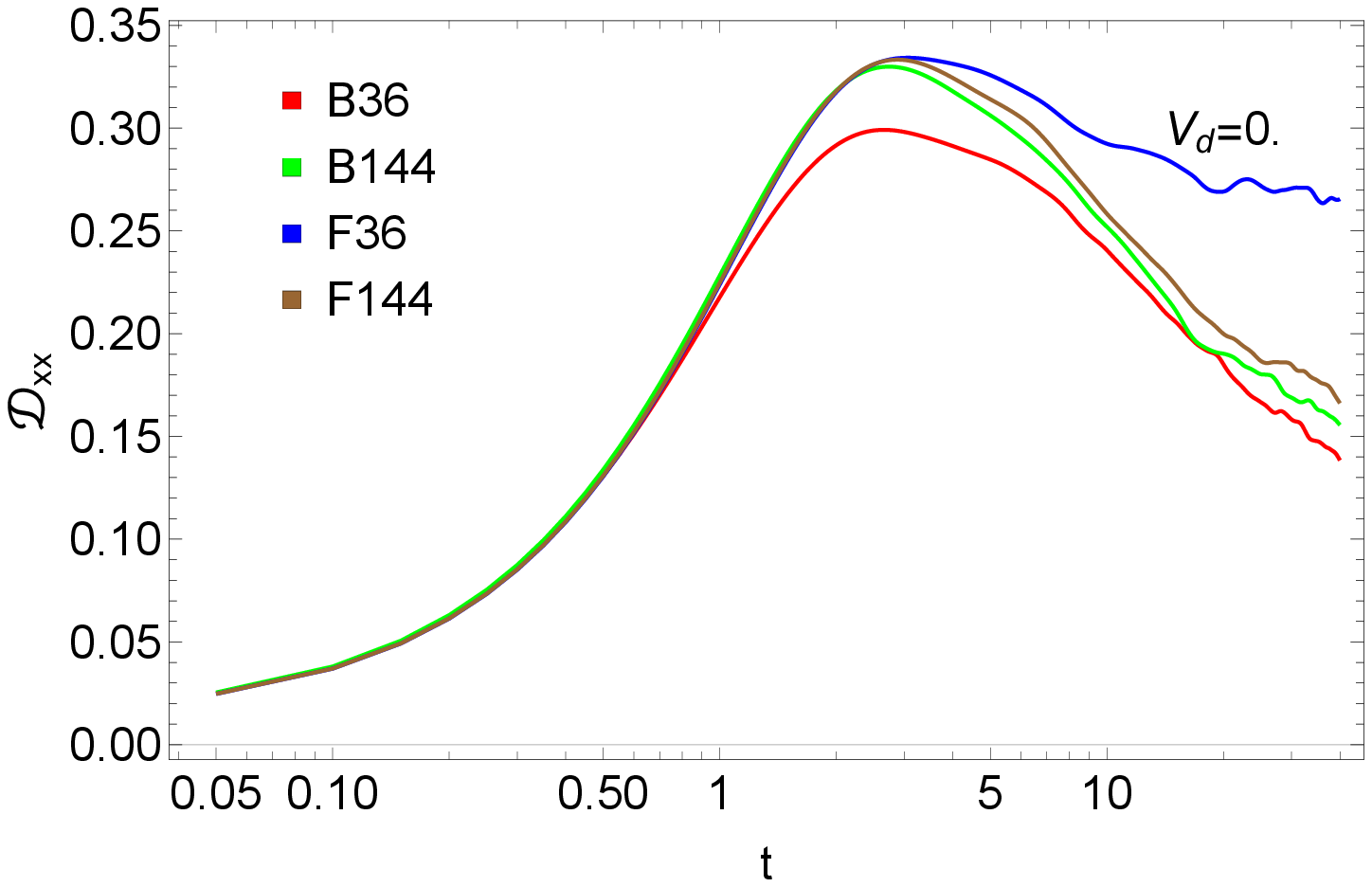}} %
	\subfloat{\includegraphics[width = 0.5\linewidth]{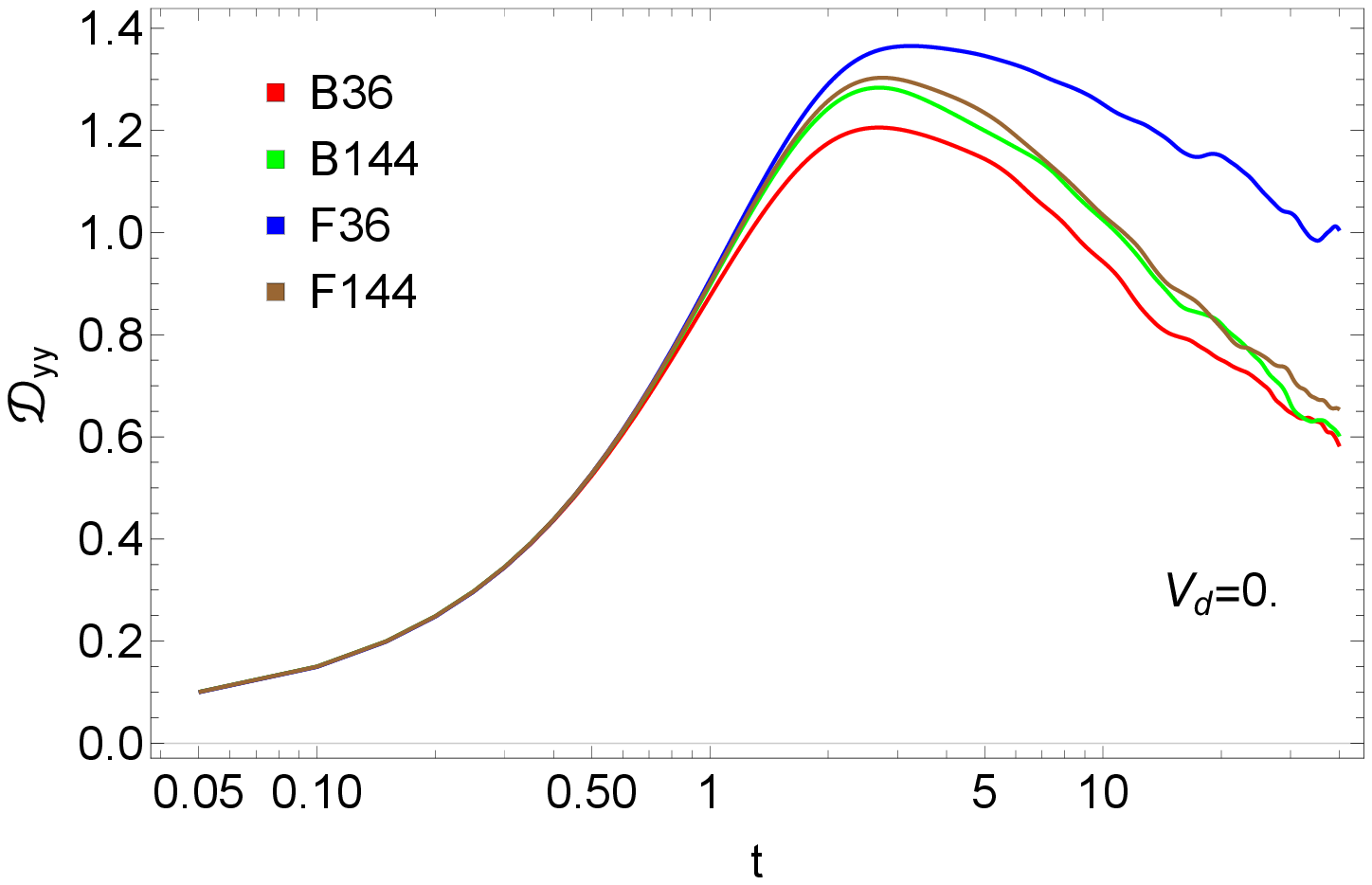}}\newline
	\subfloat{\includegraphics[width = 0.5\linewidth]{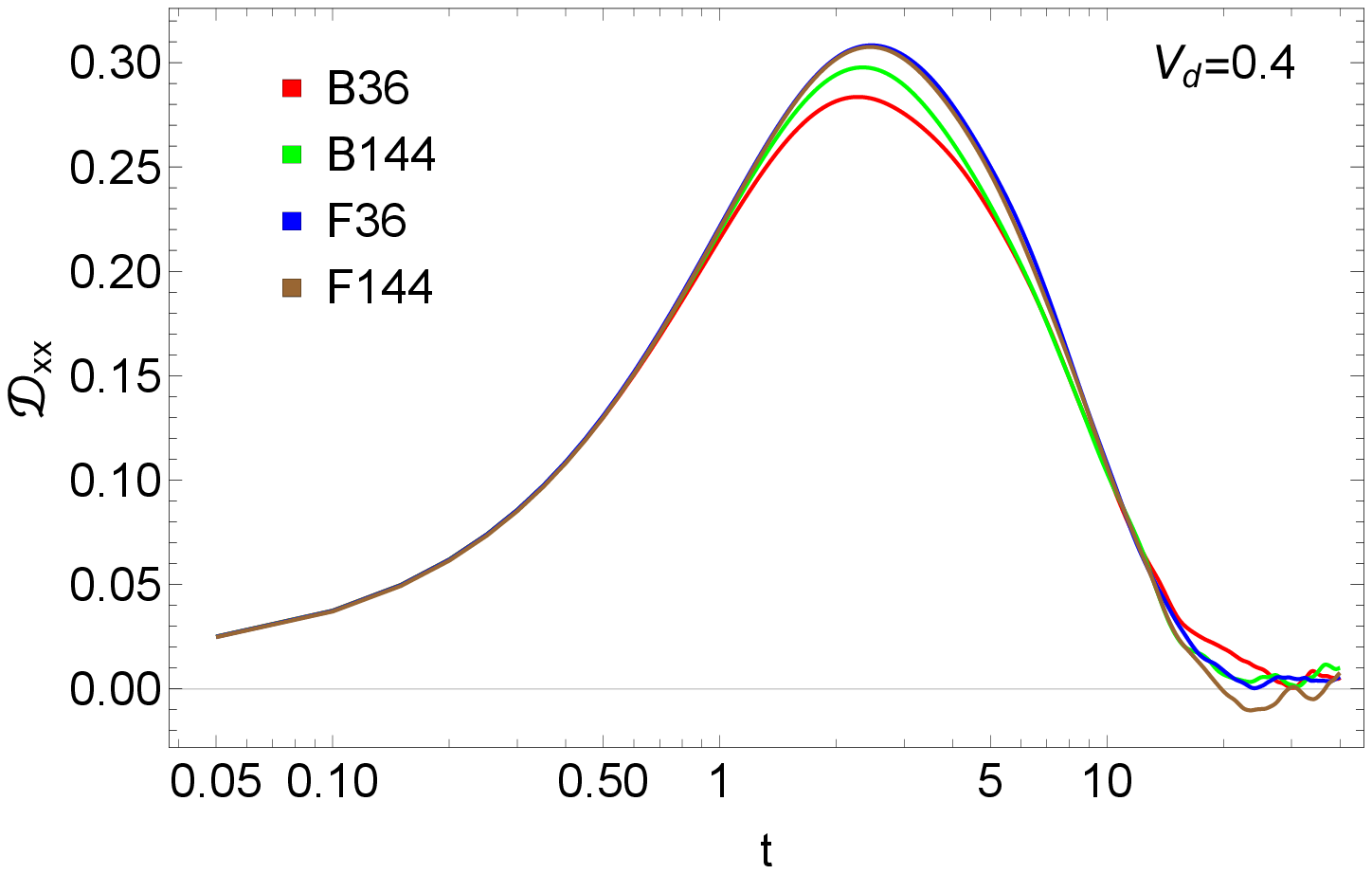}} %
	\subfloat{\includegraphics[width = 0.5\linewidth]{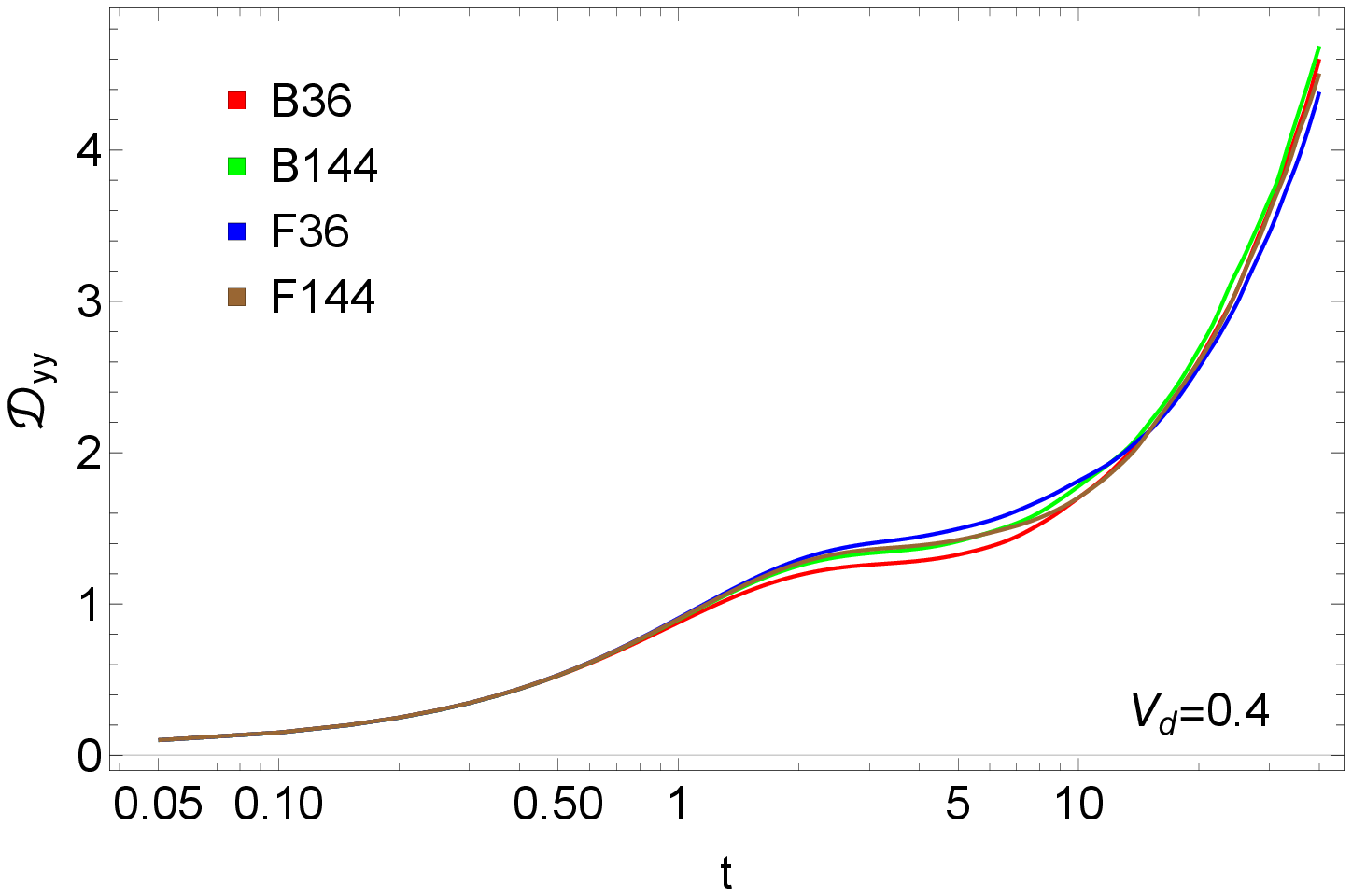}}
	\caption{Diffusion coefficients $\mathcal{D}_{ii}(t)$ obtained with the methods B36 (red), F36 (blue), B144 (green), F144 (brown).}
	\label{fig:361}
\end{figure}

The time dependent diffusion coefficients $2D_{ii}(t)=d\left\langle x_{i}^{2}(t)\right\rangle /dt$ are presented in Fig. \ref{fig:361}. The decay of $D_{ii}$ at large time is the consequence of trajectory trapping. One can see that all simulations yield practically the same result at $t<1$ and that significant differences appear at $t>1,$ especially between the calculations at $N_c=36$ and those at $N_c=144.$ This figure also shows that the $F36$ method underestimates the trapping of particles (it yields a slower decay of $D_{ii}$ at large times $t>2-3$). The converse is true for the $B36$ method which overestimates the trapping by smaller values of $D_{ii}$ at large times. All methods give a dependence of the diffusion coefficients as $\mathcal{D}\propto t^{\gamma }$. For $F36$ $\gamma \approx -0.15$ while for $B36$ $\gamma \approx -0.4$. At larger values of $N_c$, $F144$ and $B144$, we can observe how the results and the slope $\gamma $ converge towards a common profile with $\gamma \approx -0.3,$ in accordance with well known results \cite{RevModPhys.64.961}. 

\subsection{The hybrid representations}

The results from Fig. \ref{fig:361} with no bias $V_d = 0$ suggest that there are some intrinsic pathologies within the Fourier and Blob representations. The first seems to produce very long (quasi-free) trajectories, while the latter very small, closed and less complex trajectories. 
The explanation is related to the specific form of the parametric functions of each representation. Using waves (Fourier) with a small number of terms is more likely to produce long equipotential lines (which are, in fact, trajectories). A single plane wave is unable to produce a closed field line. In contrast, even a single Blob function will generate an inherently closed trajectory. A small number of Blob functions is unlikely to produce long equipotential lines. 

Also, the Table \ref{table1} suggests that the Blob method reproduces better than the Fourier method the higher-order correlations ($\mathcal{E}_{(4)}$). The overestimation of these correlations corresponds to smoother fields, which means less complex fields. 
But a less complex field has less complex equifieldlines and, consequently, less complex Lagrangian solutions. The overestimation of the correlation in the Fourier representation is natural: the waves are omnipresent, thus, any two points are "correlated" through the wave. Only the statistical averaging of the phases can decouple them.

These shortcomings do not affect the Lagrangian distribution of velocities or the average velocity of the ensemble, as seen in Figs. \ref{fig:32},\ref{fig:33}. Their effect is visible in the diffusion coefficients, which show a much stronger dependence on $N_c$. 

\begin{figure}[tbp]
	\centering
	\includegraphics[width=0.95\linewidth]{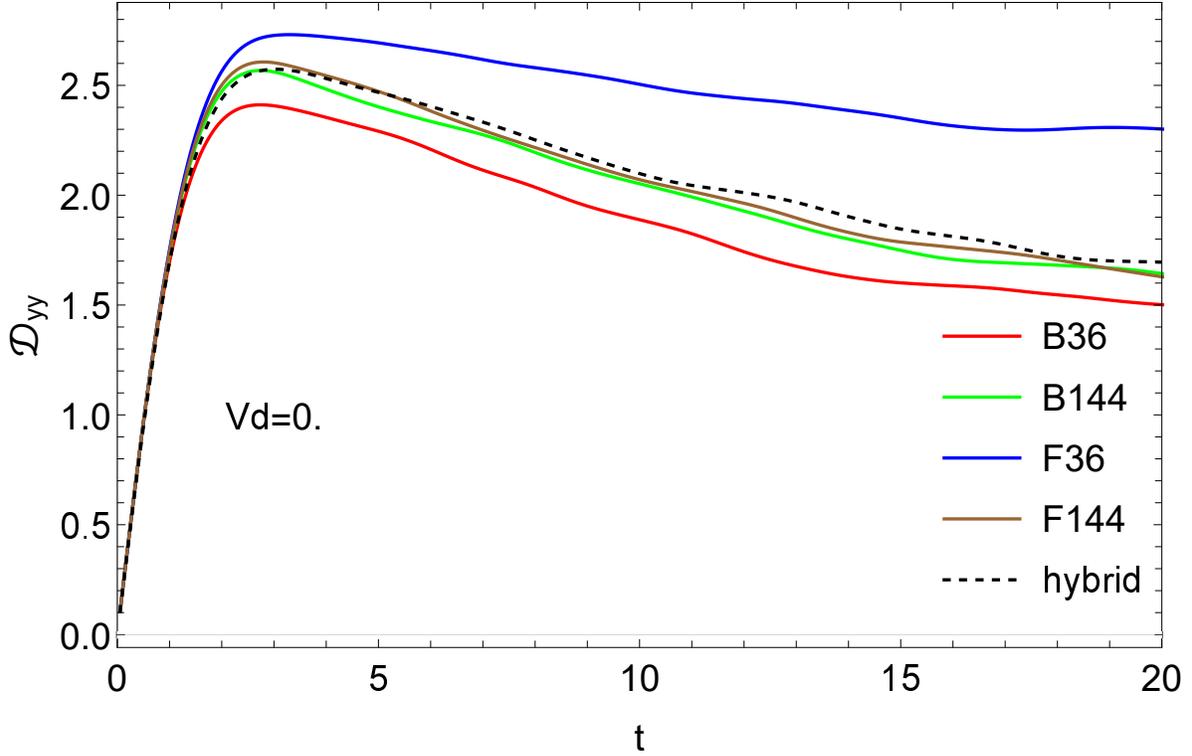}
	\caption{The diffusion coefficient $D_{yy}$ obtained with the Fourier-Blob representation \eqref{hybrid} with $N_c=36$ (for both Fourier and Blob terms) compared to the results presented in Fig. \ref{fig:361}}
	\label{fig:hib}
\end{figure}

These structural properties of the Blob and Fourier methods can be exploited to yield improved results for the diffusion coefficients without increasing $N_c$. We propose a hybrid representation of the GRF that combines the Fourier and Blob methods:

\begin{equation}
\phi_{FB}(\mathbf{x})=\eta_1\phi _{F}(\mathbf{x})+\eta_2\phi _{B}(\mathbf{x}).  \label{hybrid}
\end{equation}%
with $\eta_1^2+\eta_2^2=1$. We show that the systematic errors of the two methods compensate in this Fourier-Blob (FB) representation. The results obtained for $\eta_1=\eta_2=1/\sqrt{2}$ using the Fourier-Blob approach \eqref{hybrid}  with $N_c = 36$ are shown in Fig. \ref{fig:hib} for the diffusion coefficient $D_{yy}(t)$. The resulting profile is very close to the profiles obtained with $F144 $ and $B144$ simulations at any time.

Finally, it must be emphasized that the hybrid approach enables the possibility of reproducing accurate Lagrangian properties of stochastic transport while requiring roughly half of the CPU time required by the use of only Fourier or Blob representations. A factor of two becomes highly relevant when dealing with more complex covariance functions which may require \emph{apriori} a larger number of parametric functions $N_c$.

\section{Summary and conclusions}
\label{section_3}

The general integral representation of the GRFs \eqref{eq_1.1a},\eqref{eq_1.1b}, contains a parametric function $F(\mathbf{x};\mathbf{s})$ and an uncorrelated random variable $\zeta (\mathbf{s}).$ We have derived from Eqs.  \eqref{eq_1.1a},\eqref{eq_1.1b} a set of discrete representations. They are of Blob and Fourier type, according to the parametric function that is a space structure $F_{B}\left(\mathbf{x}-\mathbf{a}_{j}\right) $ in the first case and a wave amplitude structure in the second case $F_{F}\left( \mathbf{x};\mathbf{k}_{j}\right) =\sqrt{S(\mathbf{k}_{j})}sin(\mathbf{k}_{j}\mathbf{x})$. Additional stochastic elements were introduced in both types of representations by considering the points $\mathbf{a}_{j}$ and the wave numbers $\mathbf{k}_{j}$\ as stochastic parameters with uniform distributions. The random variable $\zeta $ was taken with discrete ($\zeta =\pm 1$) support. 

Six representations of the GRF, defined in Table \ref{table0}, were analyzed to prove that our proposal Fourier (FRD) and Blob (BRD) me provide a better convergence of the Eulerian properties than other standard representations.We have shown that reasonable errors in the covariance and in the PDF of the potential are obtained at much smaller values of $N_c$ and $M$ than in the usual Fourier representation (FFC). This leads to the decrease of the computing times by at least one order of magnitude compared to the usual FFC method.

The convergence of the Lagrangian properties of these two methods were further analyzed in the frame of the DNS of a special type of stochastic transport described by a V-Langevin equation in two-dimensional, time-independent velocity fields with zero divergence. The invariance of the Lagrangian potential in each realization and the statistical invariance of the Lagrangian velocity provide benchmarks for the validation of the numerical results. We have shown that simulations with both Fourier and Blob methods satisfy these constraints with good precision for $N_c\gtrsim 100$\ and $M\gtrsim 10^{4}$. The main difference between these representations appear in their ability to describe the effects of trajectory trapping or eddying on the contour lines of the potential.

The Fourier (FRD) results underestimate while the Blob (BRD) method overestimates the effects of trapping on the diffusion coefficients. These systematic errors were strongly reduced by a hybrid representation which combines linearly the Fourier and Blob series in a single Fourier-Blob method. The result is a representation able to decrease the value of $N_c$ required for a certain accuracy and such to reduce the calculation time by a factor $2$ compared to the BRD and FRD. 

In conclusions, we have strongly improved the representation of the GRFs by introducing additional random elements. We have shown that the hybrid Fourier-Blob method \eqref{hybrid} provides a fast tool that can be used in the numerical studies of complex stochastic advection processes. This opens the possibility of performing such  studied on personal computers. For the case analyzed here, typical running times are of the order of $10^{2}s$ or even less for the hybrid representation.

\section{Acknowledgement}

This work has been carried out within the framework of the EUROfusion Consortium and has received funding from the Euratom research and training programme 2014-2018 under grant agreement No 633053 and also from the Romanian Ministry of Research and Innovation. The views and opinions expressed herein do not necessarily reflect those of the European Commission.

\bibliographystyle{unsrt}
\bibliography{bib}
\end{document}